# Incubating advances in integrated photonics with emerging sensing and computational capabilities

Sourabh Jain,[1,2,3] May H. Hlaing,[4] Kang-Chieh Fan,[1,2] Jason Midkiff,[4] Shupeng Ning,[1,2] Chenghao Feng,[1,2] Po-Yu Hsiao,[1,2] Patrick T. Camp,[1,2] and Ray T. Chen[1,2,4,a)]

## AFFILIATIONS

[1]Department of Electrical and Computer Engineering, University of Texas at Austin, TX 78758, USA
[2]Microelectronics Research Center, The University of Texas at Austin, Austin, TX 78758, USA
[3]Centre of Excellence for Nanotechnology, Department of Electronics and Communication Engineering, Koneru Lakshmaiah Education Foundation, Vaddeswaram, Andhra Pradesh–522302, India
[4]Omega Optics, Inc., 8500 Shoal Creek Blvd., Bldg. 4, Suite 200, Austin, Texas 78757, USA

[a)]**Author to whom correspondence should be addressed:** chenrt@austin.utexas.edu

## ABSTRACT

As photonic technologies grow in multidimensional aspects, integrated photonics holds a unique position and continuously presents enormous possibilities for research communities. Applications include data centers, environmental monitoring, medical diagnosis, and highly compact communication components, with further possibilities continuously growing. Herein, we review state-of-the-art integrated photonic on-chip sensors that operate in the visible to mid-infrared wavelength region on various material platforms. Among the different materials, architectures, and technologies leading the way for on-chip sensors, we discuss the optical sensing principles that are commonly applied to biochemical and gas sensing. Our focus is on passive optical waveguides, including dispersion-engineered metamaterial-based structures, which are essential for enhancing the interaction between light and analytes in chip-scale sensors. We harness a diverse array of cutting-edge sensing technologies, heralding a revolutionary on-chip sensing paradigm. Our arsenal includes refractive-index-based sensing, plasmonics, and spectroscopy, which forge an unparalleled foundation for innovation and precision. Furthermore, we include a brief discussion of recent trends and computational concepts, incorporating Artificial Intelligence & Machine Learning (AI/ML) and deep learning approaches over the past few years to improve the qualitative and quantitative analysis of sensor measurements.

## I. INTRODUCTION TO INTEGRATED PHOTONICS

The utilization of chip-scale photonics, governing the manipulation of photons, has several unprecedented advantages over traditional electronics and larger-scale optical systems, including miniaturization and amplified performance, affording a broad spectrum of applications spanning communication systems, sensory mechanisms, imaging apparatuses, and quantum technologies.[1–4] The integration of various functional photonic devices on a chip to achieve numerous applications has become a dominant practice over the past few decades. The applications of predominant interest in this review are those that detect and quantify the presence of matter, classified here as "biosensing," which detects biological matter through its effect on the optical properties of a system (e.g., refractive index or transmission), and "spectroscopic sensing," which detects matter through its effect on the spectral properties of an optical system. Note that, in this classification scheme, a sensor may be classified as a biosensor, spectroscopic sensor, or both. The data in Fig. 1 illustrate the advancement of on-chip biosensors and on-chip spectroscopy technologies over the last two decades, indicating forthcoming progress in compact sensing technology.[5] In Fig. 1(a), the increasing attention of researchers toward biosensors can be seen, particularly the ever-increasing influence of the on-chip technique, which is a solid step toward the development of future miniaturized and portable sensors. Although Fig. 1(b) shows a decline in research on spectroscopy over the past few years, the exponential growth in on-chip spectroscopy paves the way for ultrasmall and efficient spectroscopic sensors. The compactness and scalability of integrated photonic devices to increased levels of integration and production/mass deployment render them ideal candidates for investigation and subsequent commercialization. The visible to mid-infrared region of the electromagnetic spectrum (~750–10 THz) has garnered significant attention within the scientific community for sensing purposes, leading to a number of devices that have been effectively transitioned to industrial-scale production for a wide range of sensing applications.[6,7] Several material platforms and technologies have evolved depending on the operating wavelength(s).[8,9] It is impractical to encompass all these within a single article. This review primarily incorporates recent advancements in the most prominent on-chip technologies emerging in biochemical and gas-sensing applications, mainly silicon (Si), silicon nitride (SiN), and indium phosphide (InP) material-based platforms.

Silicon exhibits minimal absorption across the wavelength range of 1.1–8.5 $\mu$m and has an exceptionally high refractive index of approximately 3.4 within this range. The most popular on-chip functions have been demonstrated on silicon-on-insulator (SOI) platforms. However, the presence of oxide in the SOI platform limits optical transparency to between 1.1 and 3.8 $\mu$m.[10] To achieve broader transparency across a range from visible to mid-infrared (approximately 300 nm to 8 $\mu$m), SiN has shown prominent optical characteristics. However, the refractive index is not as strong as that of silicon (<2) but is still satisfactory for implementing numerous on-chip functionalities. Moreover, it possesses a moderately high nonlinear refractive index ($n_2 = 2.4 \times 10^{-19}$ m$^2$/W), surpassing silica by tenfold, and crucially, it demonstrates compatibility with semiconductor mass manufacturing techniques.[11]

This review primarily focuses on advancements in miniaturized passive optical waveguides for on-chip biosensor devices and on-chip spectroscopic sensing techniques. The ultimate goal of integrated on-chip photonic sensing is to achieve a fully miniaturized sensing system, in which all components, such as the light source and detector, are seamlessly integrated with the sensing waveguide.[12–16] To address this, we also briefly discuss the integration of on-chip light sources and detectors with these compact waveguides as well as the development of integrated on-chip spectrometers. Particular attention is given to III-V material systems, especially the InP-based platform, which holds great potential for realizing fully functional lab-on-chip sensors.

Within the framework of the platforms mentioned in the preceding paragraph, regarding the technologies employed in on-chip biosensing applications, our primary focus is on advancements in refractive index sensing. Refractive index sensing encompasses a broad range of sensing techniques, including both non-plasmonic methods and plasmonic methods. Specifically, we explore cutting-edge innovations involving microring resonators and biosensors rooted in photonic crystals. Furthermore, we highlight recent strides in plasmonic-based refractive index-based sensing, considering resonances in both propagating and localized surface plasmons, as well as the popular spectroscopic technique that exploits plasmonic resonances, known as surface-enhanced Raman spectroscopy. Regarding gas sensing applications, our examination is centered on the development and achievements of on-chip absorption spectroscopy. Notably, we place particular emphasis on the mid-IR region, which has garnered substantial scientific interest because it encompasses the fundamental absorption signatures of the majority of gas molecules. We also briefly discuss recent advancements in ancillary spectroscopic techniques, including Fourier Transform Interferometer Spectroscopy, frequency comb spectroscopy, and wavelength modulation spectroscopy. Finally, we discuss the integration of artificial intelligence and machine learning techniques to enhance the analysis and predictive capabilities, followed by concluding remarks.

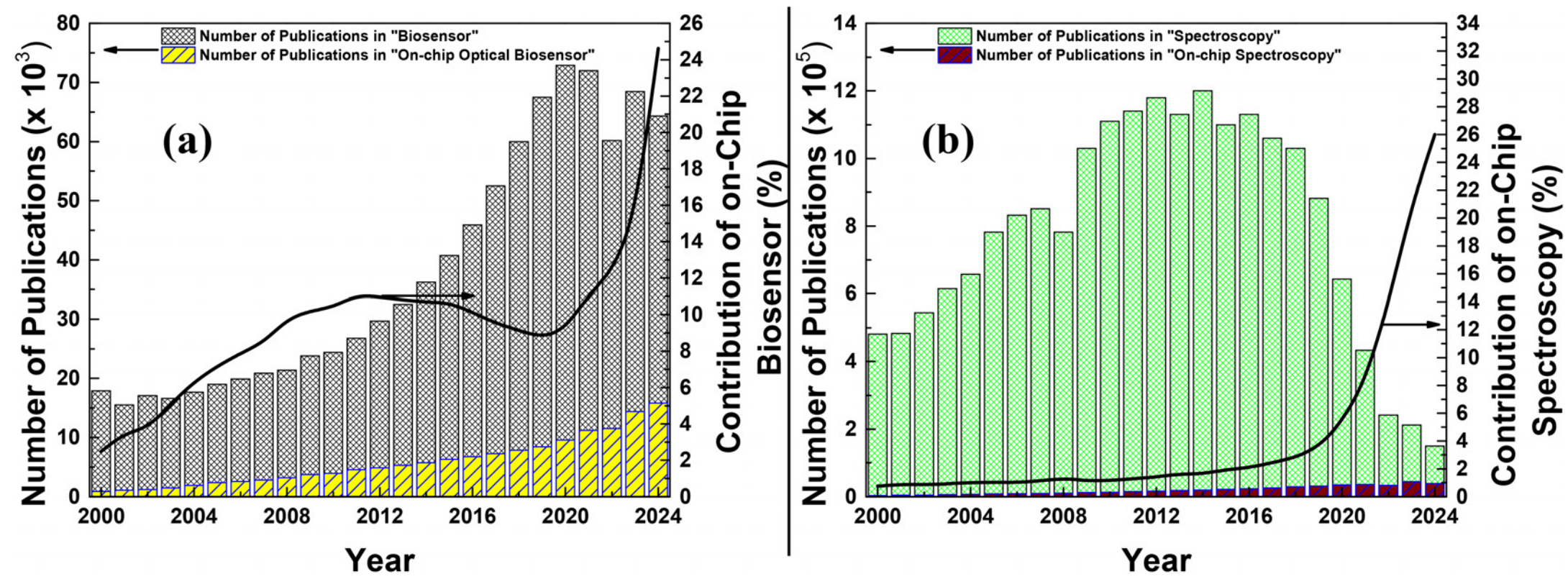

FIG. 1. Publication data obtained from Google Scholar[5] datasets over the past twenty years encompass the fields of (a) biosensors and on-chip optical biosensors and (b) spectroscopy and on-chip spectroscopy.

## II. ADVANCES IN REFRACTIVE INDEX SENSING VIA STRUCTURAL ENGINEERING APPROACHES

Refractive index is a fundamental material property that dictates the electrical, optical, and magnetic characteristics and behavior of a device. Depending on the composition, geometry, and operating wavelength, optical waves propagating in a core guiding medium surrounded by cladding experience an effective refractive index ($n_{eff}$). The chip-level manipulation of $n_{eff}$ has been extensively exploited to alter photon behavior, resulting in many exciting engineered optical properties that can be employed for various applications, including sensing and computing.[17–20] In biosensing, the change in $n_{eff}$ of the waveguide can be used to detect changes in the refractive index of the surrounding medium, often indicative of changes in the physical properties of the medium, such as the presence of biomolecules or the occurrence of biochemical reactions. Evanescent-field sensing is one of the most effective sensing methods and is most prominent at the on-chip miniaturized scale, where it can be easily incorporated using mature microfabrication processes. This technique utilizes planar and resonator waveguiding structures that permit the electromagnetic field to extend beyond the device surface, enabling interactions with analyte molecules or biomolecules nearby. When an analyte contacts the evanescent tail, it changes the $n_{eff}$ of the propagating optical mode. Information regarding targeted analytes can be extracted by tracking the fundamental properties of the propagating signal (i.e., intensity, phase, and polarization). Researchers have explored various structural geometries, the most common of which are microring resonators (MRRs), Mach–Zehnder interferometers (MZIs), photonic crystals, and plasmonic structures. Figure 2 shows the evanescent-field-sensing mechanism for various standard optical waveguide configurations.

### A. Microring resonator

Optical microring resonators are popular devices with potential applications in various fields, including telecommunications, sensing, and biomedical imaging.[25–27] The fundamental working principle of optical microring resonators involves the interference of light waves propagating around ring-shaped waveguides. Resonant coupling occurs when the circumference of the ring is an integer multiple of the wavelength of light, leading to constructive interference and a decrease in transmission at the resonant wavelength at the through-port.[28] The resonant wavelength depends on factors such as the radius of the ring, the refractive index of the material, and the number of times the light wave propagates around the ring. The resonant properties can be modified by adjusting these parameters.

In this section, we present a comprehensive overview of recent advances in optical microring resonators for sensing applications, emphasizing their performance in terms of selectivity and sensitivity. For refractive index-based optical sensors, the term sensitivity ($S$) refers to the change in the resonant wavelength shift per unit change in the analyte's refractive index, expressed as $S = \Delta\lambda_{res}/\Delta n_{analyte}$. In addition, we address the challenges and opportunities for future development. Optical microring resonators have shown promise in biosensing applications as label-free biosensors for biomolecule detection. The binding of a biomolecule to the ring's surface alters the effective refractive index ($n_{eff}$) of the device, resulting in a shift in the resonant wavelength. This shift can be quantified and correlated with the concentration of biomolecules.

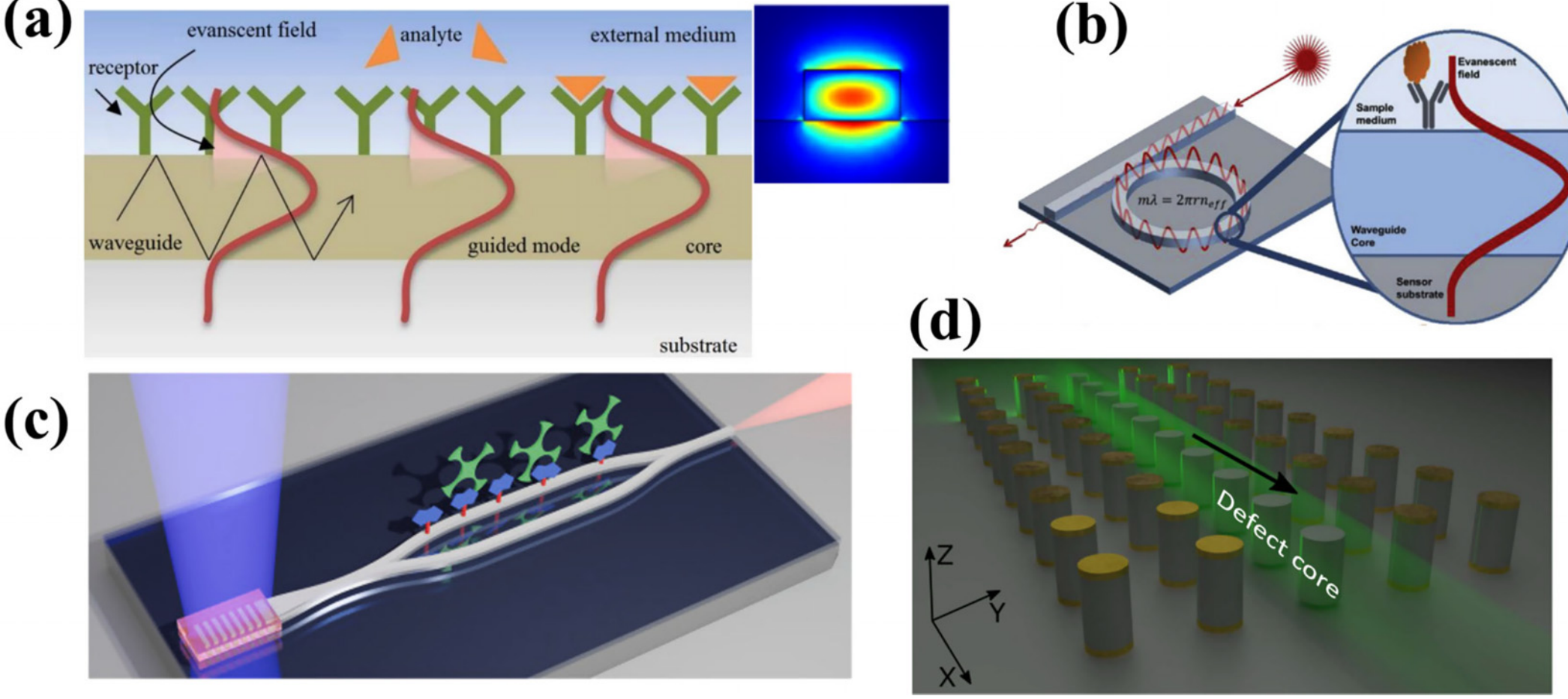

**FIG. 2.** Exemplary artistic view of optical evanescent field sensing in different optical structures. (a) Optical waveguide, accompanied by an inset displaying the optical mode profile, demonstrates the presence of an evanescent tail in the cladding (bottom) and the external medium.[21] Reprinted with permission from Estevez *et al.*, Laser & Photonics Reviews, **6**(4), 463–487 (2012). Copyright 2012 Wiley.[21] In (b), the evanescent wave extending from the microring resonator waveguide probes the local refractive index.[22] Reprinted from Cardenosa-Rubio *et al.*, Current Opinion in Environmental Science & Health, **10**, 38–46 (2019). Copyright 2019 Elsevier[22] (c) Mach–Zehnder interferometer sensing configuration in which one arm of the MZI is open to interact with the sensing molecule and another arm acts as a reference arm.[23] Reprinted from F. Vogelbacher *et al.*, Biosensors Bioelectron., **197**, 113816 (2022). Copyright 2022 Elsevier.[23] In (d), an illustration depicts a defect-mode photonic crystal waveguide situated in the central row of the waveguide.[24]

Achieving high sensitivity in simple ring-resonator-based devices (without index perturbation) on commonly used Si-, SiN-, and InP-based material platforms is challenging. This challenge arises because of the limited mode overlap of the evanescent field with the analyte, the small footprint area, background noise, and unavoidable material loss. Several key approaches can be employed to enhance the sensing performance of the ring-resonator-based on-chip devices. Optimizing the ring-resonator design for low optical losses and an exceptionally high-quality factor (Q) can result in an improved signal-to-noise ratio (SNR). For instance, Ahn *et al.* experimentally achieved an extraordinary quality factor exceeding $2 \times 10^6$ on an SOI platform by employing single-mode operation and flexible dispersion engineering, where they suppressed higher-order modes in the MRR using an inverse design approach [see Fig. 3(a)].[27] In another example, Puckett *et al.*[26] presented a $Si_3N_4$ resonator with an exceptionally high intrinsic quality factor of $422 \times 10^6$ and an absorption-limited Q of $3.4 \times 10^9$ through careful reduction of scattering and absorption losses. They demonstrated narrow resonant linewidths with an intrinsic linewidth of 453 kHz and a loaded linewidth of 906 kHz, with a remarkably low loss of 0.060 dB/m [see Figs. 3(e)–3(j)]. Noise reduction techniques, such as temperature stabilization and shielding, can increase the sensitivity while maintaining reliable device operation.[26,29–31]

Another approach involves modifying the waveguide or ring resonator surface to enhance light-matter interaction, incorporating subwavelength gratings and photonic crystal cavities to further improve sensitivity.[22,32–35] Recently, our group experimentally demonstrated a subwavelength-grating MRR for SARS-CoV-2 detection by functionalizing the sensing surface. We successfully detected and differentiated COVID-19 and influenza with a detection limit of 100 fg/ml[36] [Figs. 4(a)–4(c)]. The exceptional performance can be attributed to the optimized subwavelength grating of rectangular pillars by making up the waveguides of the microring resonators, which enhances the light-analyte interaction between the resonator and the surrounding medium. In another biosensing example, trapezoidal-shaped ring pillars were optimized, and 1 nM miRNA with anti-DNA: RNA antibody amplification was observed,[37] aided by an asymmetric effective refractive index distribution, which significantly reduced bending loss in the ring structure[38,39] [Figs. 4(d)–4(g)].

Similar to the subwavelength grating structure, incorporating photonic crystals in the ring structure can enhance light-analyte interactions, as discussed earlier. The work presented by Sun *et al.*[40] explored a one-dimensional photonic crystal microring resonator (PhCMRR) operating within a high-order photonic bandgap in the mid-IR range. The resonator demonstrated enhanced field confinement in the analyte, benefiting from its slow-light-assisted structure. A similar approach was used by Wang *et al.*[41] using two cascaded silicon-on-insulator PhCMRRs employed for dual-parameter sensing using a multiple-resonance multiple-mode technique. This approach enables simultaneous measurement of relative humidity (*RH*) and temperature using single-spectrum measurements. The obtained results included relative humidity sensitivities ($\Delta\lambda/\Delta RH$) of 3.36 and 5.57 pm/%RH, along with temperature sensitivities ($\Delta\lambda/\Delta T$) of 85.9 and 67.1 pm/°C for the selected dielectric mode and air mode, respectively [Figs. 4(h)–4(i)].[41] Subsequently, a significant enhancement in both volumetric and surface sensing sensitivity was demonstrated, with a sensitivity of approximately 248 nM/RIU utilized for the detection of deoxyribonucleic acid (DNA) and proteins at the nanomolar level, representing a more than twofold improvement compared with conventional MRRs [Figs. 4(j)–4(l)].

Table I summarizes the experimental performance of MRRs for various biosensing applications. It is evident that the incorporation of subwavelength gratings (SWG) significantly enhances sensitivity, primarily because of the expanded surface area and improved light-analyte interaction.

Among several other methods, the hybrid integration of on-chip components with nanoparticles can enhance sensitivity and local electromagnetic fields.[42–44] On-chip signal processing techniques, including filters and amplifiers, improve signal quality and reduce noise. Exploring novel material platforms, such as 2D materials[45–48] and nanocomposites, also enables new sensing functionalities.[49] Kou *et al.*[48] presented an innovative approach to detect dopamine neurotransmitters by utilizing graphene integrated into a silicon microring resonator structure. This design, incorporating Si-based waveguides and ring resonators, offers the benefit of a compact sensing surface area of 30 $\mu$m.[2] The experiments were conducted at telecom wavelengths. However, for enhanced detection of biologically significant molecules, the utilization of mid-IR wavelengths can lead to substantial advancements. Mid-IR light has longer wavelengths that allow deeper penetration into samples, improving the accuracy of refractive index determination in complex media. Additionally, it is expected to utilize the vibrational mode properties of targeted analytes along with the previously described refractive index sensing method to enhance the sensitivity and specificity for detecting low-concentration analytes. Several studies have explored the potential of waveguide-integrated graphene chemical sensors designed explicitly for mid-IR operations.[50,51]

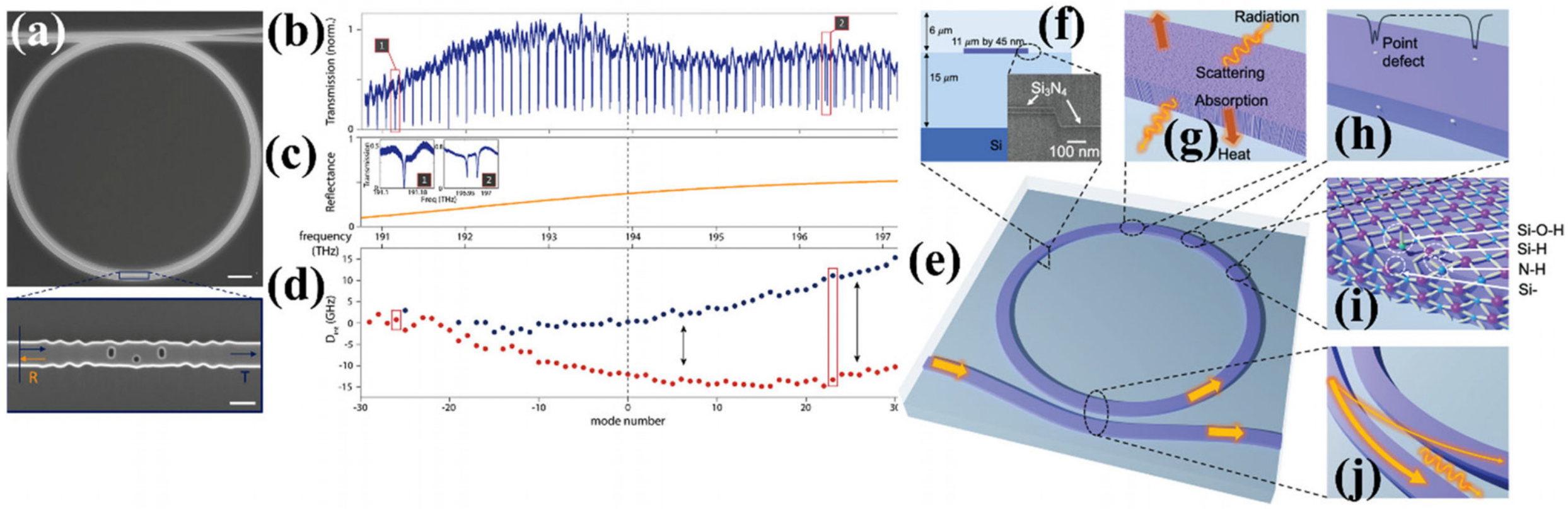

**FIG. 3.** High Q engineered ring resonator: (a) SEM image depicting a fabricated silicon microring resonator featuring a partially reflective component within the structure, (b) transmission spectrum demonstrating the presence of the inverse-designed intracavity element, (c) calculated reflectance spectrum exhibiting an increase in reflectance as the frequency rises, (d) the integrated dispersion of the split fundamental mode illustration, with blue and red points denoting each side of the split modes.[27] (a)–(d) reprinted with permission from G. H. Ahn *et al.*, ACS Photonics **9**(6), 1875–1881 (2022). Copyright 2022 American Chemical Society.[27] (e) Illustration of ultrahigh-Q MRR. (f) The cross section exhibits bottom cladding, a silicon nitride waveguide core created through etched low-pressure CVD and an additional thin blanket layer over the entire surface. (g) Scattering via the rough surface of the waveguide causes radiation mode, whereas bulk absorption results in heat. (h) The resonance experiences splitting owing to point defects. (i) Surface absorption occurs because of defects, such as Si-O-H, Si-H, N-H, and dangling bonds. (j) The bus ring coupler serves the dual function of dispersing energy among the radiation modes and inducing additional loss.[26] (e)–(j) Reprinted with permission from M. W. Puckett *et al.*, Nat. Commun. **12**(1), 934 (2021). Copyright 2021 Authors, licensed under a Creative Commons Attribution (CC BY) license.[26]

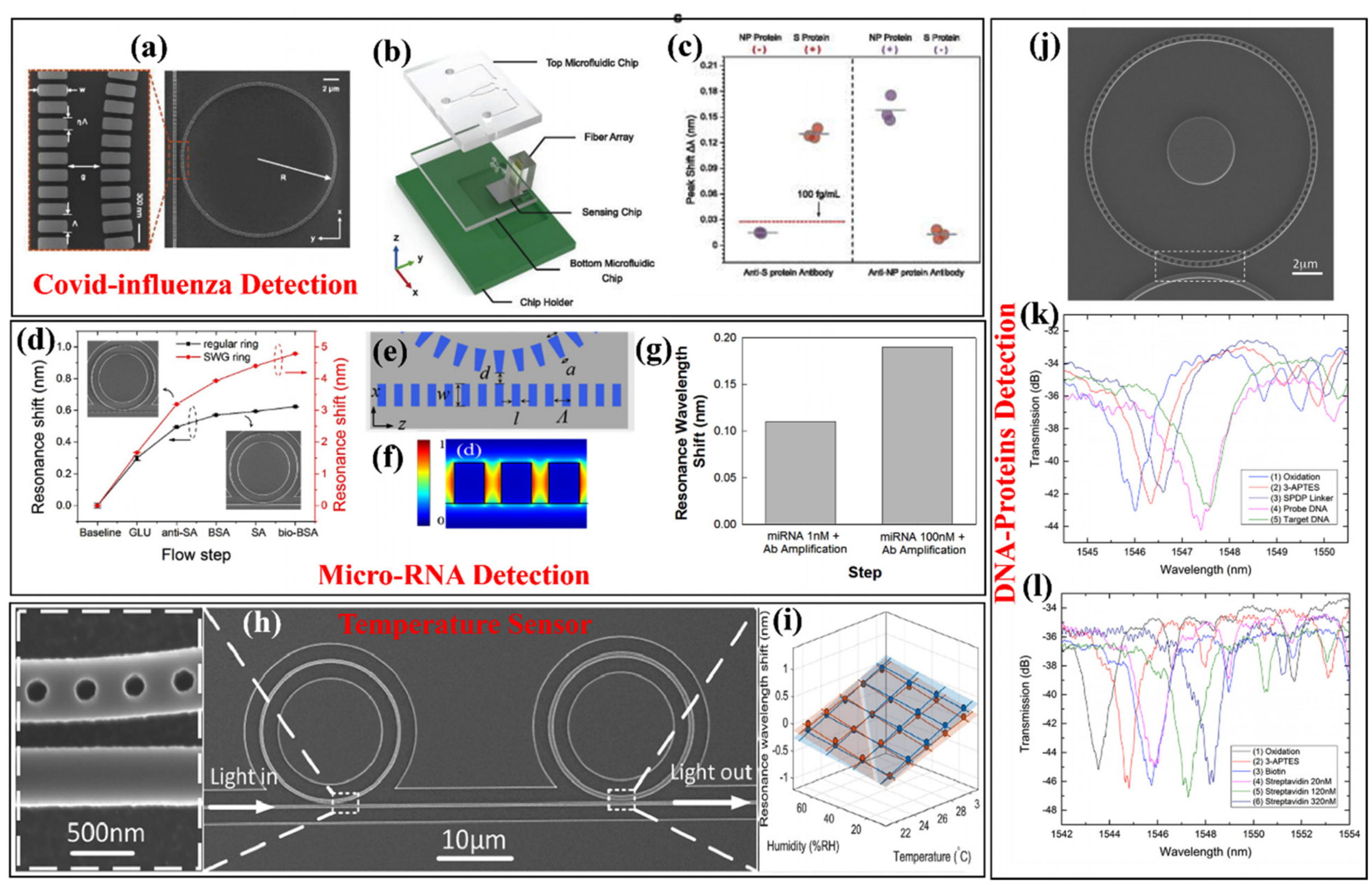

**FIG. 4.** Numerous applications were experimentally demonstrated on index-perturbed microring resonator platforms. (a) SEM image of subwavelength grating microring resonator. (b) Virtual representation of the exploded biosensing platform. (c) Testing for cross-reactivity between SARS-CoV-2 and influenza.[36] Reproduced from S. Ning *et al.*, **10**(2), 2023, Applied Physics Reviews with the permission of AIP Publishing.[36] (d) Shift in resonance of regular and SWGMRRs during the sequence of layer functionalization to detect micro-RNA. (e) Trapezoidal-shaped ring schematic and electric field distribution of the optical mode in the bus waveguide (f). (g) Experimental validation of micro-RNA on chip.[37] (h) SEM images of two cascaded photonic crystal microring resonators. (i) Measured and predicted resonant wavelength shifts under different humidity and temperature conditions.[41] (h and i) reprinted with permission from J. Wang *et al.*, Opt. Express **30**(20), 35608–35623 (2022). Copyright 2022 The Optical Society.[41] (j) SEM image of photonic crystal microring resonator. (k) and (l) Experimental demonstration of DNA and protein detection, respectively.[35] (j)–(l) Reprinted with permission from S. M. Lo *et al.*, Opt. Express **25**(6), 7046–7054 (2017). Copyright 2017 The Optical Society.[35]

**TABLE I.** Comparative analysis of experimental MRR performance across various biosensing applications.

| MRR type | Q-Factor | Sensitivity | Application |
|---|---|---|---|
| Strip Waveguide | 20 000 | 70 nm/RIU | Protein detection[52] |
| Ultra-thin Rib Waveguide | 25 000 | 100 nm/RIU | Temperature detection[53] |
| Rectangular SWG | 1200 | 248 nm/RIU | DNA and protein[35] |
| Rectangular SWG | 2180 | 423 nm/RIU | Streptavidin[54] |
| Rectangular SWG | 1650 | 437.2 nm/RIU | SARS-CoV-2[36] |
| Trapezoidal SWG | 9100 | 440.5 nm/RIU | miRNA[37] |

### B. Photonic crystal-based biosensor

Photonic crystal waveguides (PCWs) have emerged as pioneering breakthroughs in photonics, offering distinctive characteristics for various applications. These waveguides manipulate the light propagation by introducing periodic variations in the refractive index within the dielectric structure. Bragg reflection and defect-band-guided propagation are two distinct phenomena that occur in photonic crystal waveguides. Bragg reflection relies on the periodic variation of the refractive index in the waveguide, causing wave reflection of specific wavelengths while allowing others to pass through. Defect-band-guided propagation occurs when a line defect (in real space) leads to a propagating mode that would otherwise be forbidden (in wavevector space) by a photonic bandgap (PBG) at the frequency of interest. Similar to the microring resonators discussed earlier, evanescent field sensing can also be utilized on a platform of photonic crystal waveguides utilizing both phenomena. Additionally, through bandgap engineering techniques, dispersion-induced slow light can be leveraged to enhance absorption spectroscopy, as detailed in the corresponding section (Sec. IV A).

At the chip scale, PCWs can be based on one-dimensional (1D)[61–64] or two-dimensional (2D)[65] photonic crystals. In the case of a 1D PCW, the mode is completely index-guided, and the crystal is used to control the dispersion characteristics. For a 2D PCW, the mode is still index-guided in the out-of-plane direction but is typically PBG-guided within the plane of propagation, and the defect band imparts dispersion characteristics. In either case, the large overlap of the mode with the sensing medium engenders pronounced interactions with the analytes, and an exceptional level of sensitivity to the surrounding environment is achieved by incorporating functional materials or by employing surface modifications to the waveguide. As such, these devices have exhibited remarkable sensitivity for the detection of biomolecules, such as DNA, proteins, and chemical markers.[55–59]

Creating microcavities from finite-length PCWs, Yang *et al.*[60] identified three distinct biomarkers in the plasma of individuals with pancreatic cancer [Figs. 5(a)–5(d)]. Microcavity PCWs were formed from 13-modified-hole line defects. Detection was accomplished with microcavities lacking nanoholes [Fig. 5(a)] and with an increased sensitivity to microcavities featuring nanoholes [Fig. 5(b)]. Remarkably, the experimental setup detected a concentration as low as 8.8 femtomolar (334 fg/mL) of the targeted biomarker in the plasma samples of the patient, which corresponds to a 50-fold dilution compared with ELISA. Despite the extraordinary engineered optical behavior at a miniaturized chip scale, PCW-based structures are sometimes incompatible with other integrated optical components because of the precise fabrication processes required and their unique optical properties, which can lead to inefficient coupling and complicated mass-manufacturing processes.[66,67] Torrijos *et al.*[61] addressed the incompatibility problem by utilizing 1-D PCW. They reported the first instance of slow-light bimodal interferometric behavior within an integrated 1D-PCW and demonstrated ethanol sensing [Figs. 5(e)–5(i)]. Table II presents the empirical findings on the performance of biosensors across a spectrum of applications. Of particular note is the relatively limited number of experimental demonstrations in the realm of 2D-PCWs, despite the wealth of theoretical research highlighting their potential utility in biosensing. The lack of experimental validation is primarily due to the complex and fragile attributes of 2D-PCWs. These characteristics impose strict limitations on the fabrication flexibility and make the device features highly sensitive to minor variations in the fabrication process. Conversely, there has been a recent surge in interest in the 1D-PCW. Notably, deliberate engineering of multimodal interference via dispersion-engineered phenomena holds significant promise for the development of high-performance biosensors for future research and innovation.

Finally, we note that, unlike biochemical applications for which refractive-index sensing is well-suited, this is generally not the case for gas sensing applications, primarily because of the limited sensitivity and selectivity to gas molecules, as gas molecules produce a negligible change in refractive index compared to other fluid analytes and the surrounding air cladding. However, plasmonic structures can enhance the interaction between light and gas analytes and thus have the potential to address the issues faced by conventional refractive index sensors, as discussed in Sec. III.

## III. ADVANCES AND APPLICATIONS OF PLASMONIC METHODS IN REFRACTIVE INDEX DETECTION AND SURFACE ENHANCED RAMAN SPECTROSCOPY

Plasmonics is a field of study that focuses on the interaction between light and free electrons in metals, specifically the collective oscillations of these electrons known as plasmons.[70–72] Plasmonic phenomena occur when light interacts with nanoscale metallic structures, enabling the manipulation and confinement of light on the subwavelength scale. This interaction is most pronounced in the visible to near-infrared (NIR) range, where the frequency of light closely matches the natural frequency of plasmonic oscillations in metals, such as gold and silver, resulting in efficient energy transfer and strong plasmonic resonance. This unique effect can be leveraged for high-sensitivity refractive index sensing and paves the way for unique on-chip sensing applications. However, plasmonic effects can also occur in the mid-infrared (mid-IR) range, where they are particularly useful for applications such as chemical sensing, despite the resonance being less pronounced owing to the lower energy of the mid-IR light. We discuss different types of plasmonic-based sensors in subsequent subsections.

### A. Surface plasmon resonance sensor

Surface plasmon resonance (SPR) sensors are advanced optical devices employed for highly sensitive molecular interaction analyses on sensor chips.[73,74] These sensors exploit the surface plasmon resonance phenomenon, which involves the collective oscillation of free electrons at the interface of a metal (typically gold or silver) and dielectric medium. Surface plasmons are associated with an electromagnetic field in the dielectric, forming a surface plasmon polariton (SPP) (in the case of a planar interface). SPPs are excited by directing a light beam onto this interface at a specific resonance angle, resulting in a distinct dip in the reflected light intensity. This dip, which is extremely responsive to changes in the refractive index of the dielectric near the metal surface, serves as a reliable indicator of the molecular binding or dissociation events in the vicinity. Immobilizing one binding partner on a metal surface enables real-time monitoring of molecular interactions by measuring shifts in the resonance angle or reflected light intensity. SPR sensors have diverse applications in fields such as biochemistry, pharmaceuticals, and environmental monitoring, offering label-free detection, real-time analysis, and multi-interaction capabilities.[75–77]

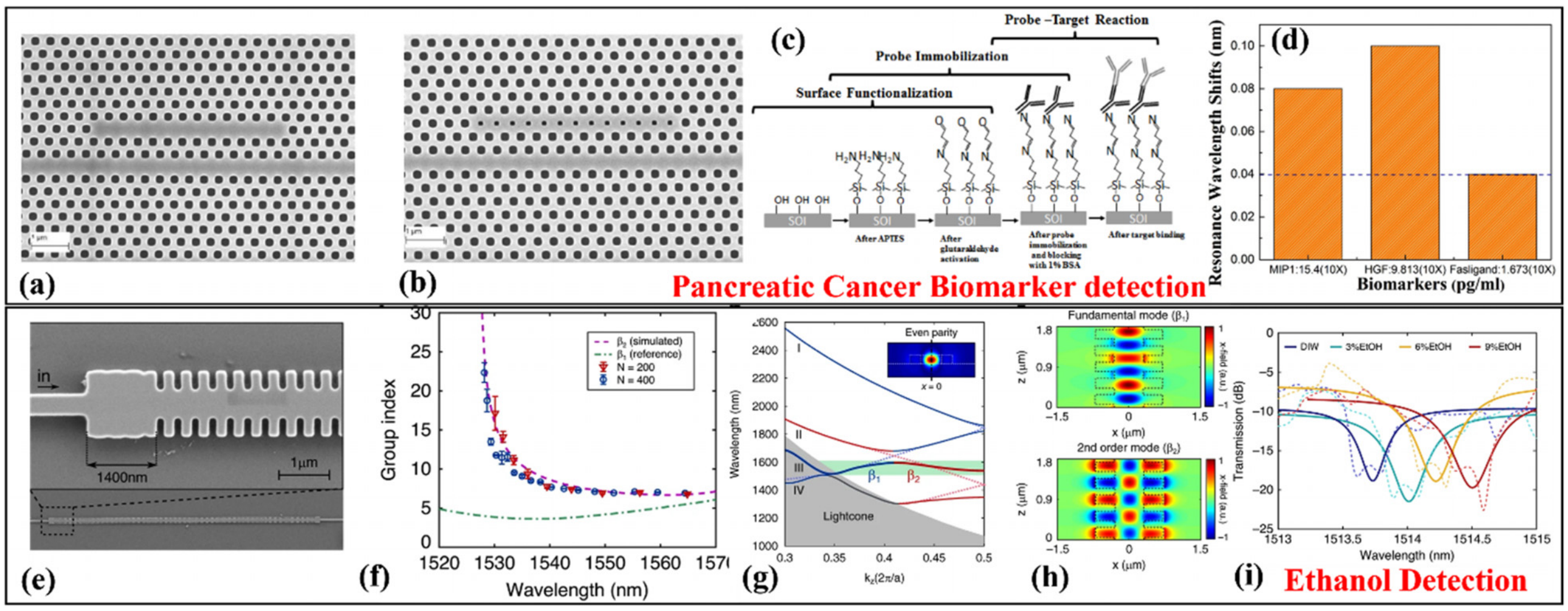

FIG. 5. Exemplary photonic crystal biosensors from published studies: (a) SEM images of an L13 PC microcavity, (b) L13 PC microcavity with nanoholes, (c) chemical structure involved in the biosensing reaction for detection of pancreatic patient plasma, (d) resonance shift observed for three biomarkers (Fasligand, MIP1, HGF).[60] (e) SEM image of a 1D photonic crystal waveguide, focusing on single-mode and bimodal sections; (f) Experimental group index data were acquired from interference points, utilizing the simulated fundamental mode group index as a reference; (g) a dispersion diagram illustrating the first three x-even parity bands for TE-like polarization, (h) real part of the electric field's x-component for both the fundamental and higher-order modes; (i) experimental optical spectra with different concentrations of ethanol in deionized Water.[61] (e)–(i) Reprinted with permission from L. Morán *et al.*, Light **10**(1), 16 (2021). Copyright 2021 Authors, licensed under a Creative Commons Attribution (CC BY) license.[61]

Chip-based SPR sensors offer several advantages, including label-free detection, dynamic measurement of binding kinetics, and high sensitivity, making them more efficient and versatile compared to traditional biosensing methods that often require complex sample preparation or labeling. However, standard SPR chips have limitations, such as restricted polarization compatibility, low selectivity, and shallow penetration depth in the sensing analyte region. Modifications such as long-range SPR chips that provide longer interaction lengths have been developed to overcome these challenges.[78,79] These advancements have enhanced the capabilities of chip-based SPR sensors and expanded their potential for diverse applications. SPR excitation requires a coupling medium to impart photon momentum at the interface, which can be accomplished using high-index prisms, gratings, waveguides, or optical fibers.[75] Grating-coupled structures have been introduced into SPR on-chip devices to enable their miniaturization and integration into lab-on-a-chip systems. However, grating-coupled SPR systems face limitations owing to their lower sensitivity compared to prism-coupled systems. To overcome this issue, Rossi *et al.*[80] integrated sensing areas with gratings in microfluidic chambers and achieved a 30%–50% increase in the overall sensitivity, making the grating coupler more competitive. The SPR on-chip is extensively used in biomedical research for studying protein-protein interactions,[81] DNA hybridization,[82] and ligand-receptor binding.[83] SPR-based biosensors are also popular for clinical diagnostics to detect and quantify biomarkers related to various diseases, such as cancer,[84–86] infectious diseases, viruses,[87] and cardiovascular disorders.[88] Huang *et al.*[89] developed a cost-effective and efficient SPR device to sense the SARS virus in one shot. The work was performed in the visible range near a 640 nm resonant peak, enhancing the sensitivity of the device. By combining a plasmonic biosensor with a standard 96-well plate or chip cartridge, the device achieved rapid detection (<15 min) with high sensitivity (limit of detection = 370 viral particles/mL), directly detecting the entire virus [Figs. 6(a)–6(c)]. In another approach to COVID-19 detection, Basso *et al.*[92] successfully detected SARS-CoV-2 antibodies in the serum of patients using

TABLE II. Comparison of the experimental performance of photonic crystal waveguides for diverse biosensing applications.

| Device type | Q-Factor | Sensitivity | Application |
|---|---|---|---|
| L13 defect hexagonal 2D PCW | 22 000 | 112 nm/RIU | Pancreatic cancer biomarker detection[60] |
| L3 defect hexagonal 2D PCW | 26 760 | 15 ng/ml | Antibodies detection[68] |
| Slotted hexagonal photonic crystal | 4000 | 500 nm/RIU | Avidin/biotin detection[65] |
| Line defect hexagonal 2D PCW | ⋯ | 94.2 pm/$\mu$M | DNA detection[69] |
| 1D Biomodal PCW | ⋯ | 150.83 $2\pi$ rad/RIU | Ethanol detection[61] |
| 1D Asymmetric multimode PCW | 861 | 325 nm/RIU | NaCl detection[62] |

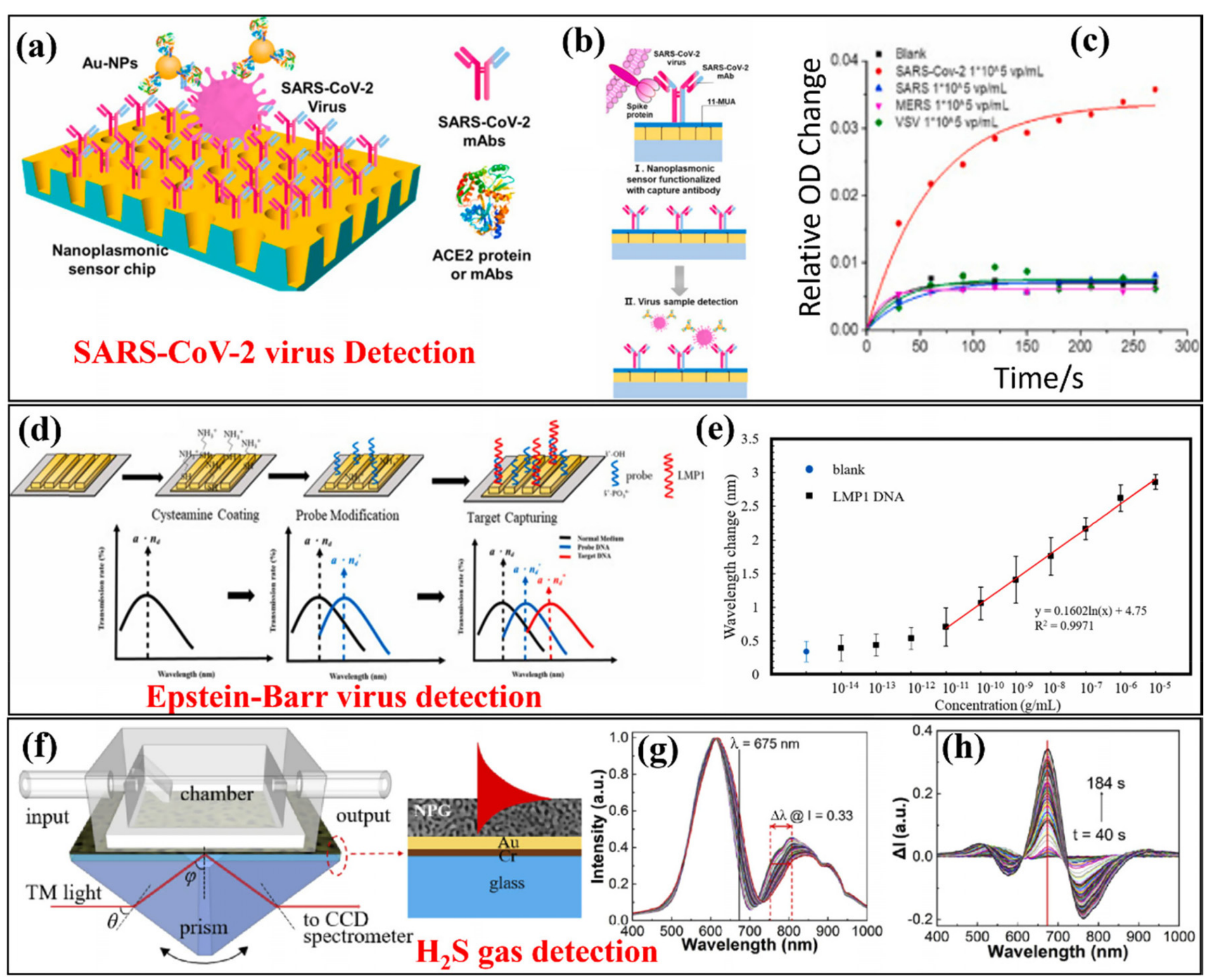

**FIG. 6.** State-of-the-art experimental demonstrations utilizing surface plasmon resonance. (a) Schematic of the plasmonic nano-cup array sensor. (b) The diagram illustrates the interaction between the spike protein situated on the surface of SARS-CoV-2 and the respective monoclonal antibodies targeting SARS-CoV-2. (c) Testing the specificity of AuNP-enhanced virus detection: Dynamic binding curves showing changes in optical density (OD) of SARS-CoV-2 antibodies with various SARS-CoV-2, SARS, MERS, and VSV pseudo-viruses at a concentration of $1.0 \times 10^5$ vp/ml.[89] (a)–(c) Reprinted with permission from L. Huang *et al.*, Biosensors and Bioelectronics **171**, 112685 (2021). Copyright 2021 Authors, licensed under a Creative Commons Attribution (CC BY) license.[89] (d) Au-capped nano-slit SPR chip layer functionalization with a probe for DNA detection, and (e) the calibration curve for latent membrane protein 1 DNA sensing is presented using the continuous microfluidic PCR-integrated nanoslit-SPR sensor.[90] (d-e) Reprinted with permission from H. Y. Hsieh *et al.*, Biosensors and Bioelectronics **195**, 113672 (2022). Copyright 2022 Authors, licensed under a Creative Commons Attribution (CC BY) license)[90] (f) Depiction of Kretschmann-type spectral NPG-SPR sensor platform for in situ gas detection at room temperature. (g) and (h) Spectra of reflected light intensity recorded at various time points following the introduction of 100 ppm $H_2S$ gas and diverse spectral variations compared with the resonance spectrum captured at 40 s.[91] (f)–(h) reprinted with permission from C. Liu *et al.,* Opt. Lett. **47**(16), 4155–4158 (2022). Copyright 2022 The Optical Society.[91]

Au-SPR chips functionalized with self-assembled monolayers (SAMs) composed of 11-mercaptoundecanoic acid (MUA). The SAMs were used to immobilize SARS-CoV-2 spike (S) and nucleocapsid (N) proteins on the gold surface, enabling the detection of antibody binding. Recently, Hsieh *et al.*[90] introduced a practical approach that integrated microfluidic polymerase chain reaction (PCR) with an SPR sensor based on gold nanoshells. This method was used to detect the DNA sequence associated with the latent membrane protein 1 (LMP1) DNA [Figs. 6(d)–6(e)]. In addition to biosensing applications, SPR chip-based platforms have attracted considerable interest for gas-sensing applications.[93–95] Zhang *et al.*[91] demonstrated gas sensing using nanoporous gold (NPG) thin films to increase the surface area for molecular adsorption. The material was tested for its response to toxic $H_2S$ gas at room temperature. By adjusting the dealloying time to control film porosity, the sensor achieved its optimal sensitivity to $H_2S$. Compared to a conventional Au-SPR sensor, the NPG-SPR sensor showed at least six times higher sensitivity to 100 ppm $H_2S$ [Figs. 6(f)–6(h)].

## B. LSPR-based sensor

Localized surface plasmon resonance (LSPR), a subtype of surface plasmon resonance, shows promise for use in highly sensitive plasmonic biosensing. LSPR is characterized by the confinement of electronic oscillations within a metallic nanostructure associated with greatly enhanced electric fields at its surface, allowing the detection of changes in the refractive index of the surrounding medium within a short distance of a few tens of nanometers. Scattering and absorption effects dominate when colloidal nanoparticles exhibit random orientation. Mie theory offers a framework for analyzing the extinction cross section (incorporating absorption and scattering) of a metallic nanosphere characterized by radius $R$ and dielectric constant ($\epsilon_m$), surrounded by a dielectric material ($\epsilon_d$).

$$\sigma_{ext} = 12\,\frac{w}{c}\,\pi\epsilon_d^{3/2}R^3\,\frac{Im(\epsilon_m)}{[Re(\epsilon_m)+2\epsilon_d]^2+[Im(\epsilon_m)]^2}. \quad (1)$$

The extinction cross section ($\sigma_{ext}$), which incorporates both absorption and scattering, depends on the nanoparticle's radius and the dielectric properties of both the metal and the surrounding medium. Typically, LSPR sensors are fabricated by constructing metallic nanostructures such as nanospheres, nanorods, nanoshells, nanowires, and nanoprisms, along with an additional sensing film layer. The efficiency of electrolysis enhancement and light quenching in LSPR is significantly influenced by the proximity, size, shape, and material of nanoparticles. Therefore, precise control over the size and shape of nanoparticles is necessary to achieve the desired optical response.[96,97] Progress in nanolithography techniques has enabled accurate fabrication of such nanostructures on standard substrates, thereby enabling the use of LSPR-based sensing at the chip scale. These chip-based substrates offer high sensitivity, repeatability, and the capability to function with other sensing systems, such as microfluidics.[87,98]

LSPR is used for transmission measurements, whereas SPR identifies changes in reflectance upon the attachment of a material to a single particle.[99,100] The main benefit of LSPR lies in the simplified setup of the measurement system, which eliminates the need for movable components in reflection-angle measurements. The transmission signal response has been utilized in various biomolecule detection scenarios by monitoring the plasmon peak shift in nanoparticle sensors as opposed to reflective signal sensors based on LSPR.[101]

Wang *et al.*[105] developed a novel biosensor utilizing gold nanorods to harness the benefits of LSPR. The biosensor was specifically designed for the detection of the hepatitis B surface antigen, which serves as an indicator of active viral replication in the hepatitis B virus. Recently, Sharifi *et al.*[102] designed an ultrasensitive LSPR biosensor using gold crystalline nanostructures to detect CA-19–9 protein in serum samples from patients with pancreatic cancer [Figs. 7(a)–7(c)]. Through meticulous optimization, the biosensor achieved a detection limit of 0.0001 U/mL. Qiu *et al.*[103] used gold nanoparticles coated with complementary DNA to validate the application of LSPR sensing for detecting oligonucleotides encoding specific sequences of SARS-CoV-2. They achieved this by integrating the LSPR technique with a thermo-plasmonic effect, successfully attaining an exceptional sub-picomolar (sub-pM) detection limit [Figs. 7(d)–7(f)]. In a recent study, mid-IR ozone gas detection was reported by Ghanim *et al.*,[106] in $SiO_2$ film utilizing LSPR enhanced micrometer-sized antenna [Figs. 7(g)–7(k)]. The sensitivity of LSPR sensors can be improved by optimizing the nanoparticle size and shape for enhanced plasmonic resonance and by enhancing surface functionalization techniques to increase the binding affinity of the target molecule.[107]

## C. Surface-enhanced Raman spectroscopy

Surface-Enhanced Raman Spectroscopy (SERS) is an advanced analytical method that amplifies weak Raman scattering signals, which are the inelastic scattering of photons by molecules that provides unique vibrational information about their chemical structure. SERS enhances these weak signals by exploiting their interactions with metallic nanostructures. Incident laser excitation triggers the emergence of localized surface plasmons with associated enhanced electric fields near nanostructures. Consequently, Raman scattering signals emitted by molecules in the vicinity of nanostructures undergo pronounced enhancement, enabling the detection and characterization of molecules, even at minute concentrations.[108,109]

The origins of SERS can be attributed to pioneering investigations conducted in the 1970s by Fleischmann, Hendra, and McQuillan,[110] as well as Albrecht and Creighton,[111] wherein they documented the observation of heightened Raman scattering signals from pyridine adsorbed on uneven metal surfaces. Subsequently, substantial advancements have been made in understanding the fundamental principles governing SERS and the development of metallic nanostructures tailored to yield superior enhancement factors. Two primary mechanisms contribute to SERS enhancement: electromagnetic enhancement and charge transfer (also termed chemical enhancement, which is a non-plasmonic condition).[112] Electromagnetic enhancement can be viewed as an extreme scenario of surface-enhanced fluorescence (SEF), wherein the limited Raman scattering cross section. inhibits plasmon quenching, a process in which the energy from excited surface plasmons is dissipated as heat or absorbed by the material, reducing the overall signal intensity instead of enhancing it. Consequently, the magnitude of the electric field in the Raman signal became insignificant. However, there is a localized field enhancement due to the excitation of plasmons, which is 108 times the normal Raman scattering, thus offering an overall enhancement.[113,114]

Silver and gold nanomaterials have emerged as highly effective options for boosting SERS signals and enhancing the detection sensitivity. They provide rapid, efficient, cost-effective, and dependable SERS platforms for detecting even minute quantities of biomolecules and pathogens down to the single-cell level. This is because of their distinct chemical, physical, and optical properties. Notably, core/shell structures such as Au/Ag nanoparticles exhibit exceptional performance as SERS substrates owing to their significantly higher enhancement factors (EFs).[115–117] Vancomycin-coated SERS facilitates the swift, culture-free detection of bacteria in clinical samples, offering a 1000-fold increase in capture efficiency without introducing significant spectral interference from the coating or other components in complex samples, such as blood. Unlike other functionalization agents such as antibodies, which can produce sharp peaks and broad background signals, vancomycin coating introduces minimal interference, enabling selective bacterial isolation from human blood. Additionally, it holds promise for drug resistance testing, thus advancing SERS-based biochips for clinical micro-organism detection.[118]

Bai *et al.*[119] presented a novel SERS platform called liquid interface-assisted SERS (LI-SERS). It enables the label-free trace detection of biomolecules, with detection limits ranging from pM to fM. The LI-SERS method demonstrated ultrahigh sensitivity and versatility in detecting DNA and $\beta$-Amyloid ($A\beta$) in trace concentrations, featuring its unique capabilities for analytical applications [Figs. 8(a)–8(h)]. In another study, Hu *et al.*[120] engineered a high-performance, homogeneous SERS chip by integrating Au/Ag nanorods with a carboxymethylcellulose hydrogel, achieving high sensitivity in trace-level pesticide residue detection. The study exhibited excellent performance in detecting thiram residues in fruits, including those with low and abundant pigment interferents such as methanol, apples, and blueberries, with detection limits of 42, 58, and 78 ppb, respectively [Figs. 8(i)–8(o)].

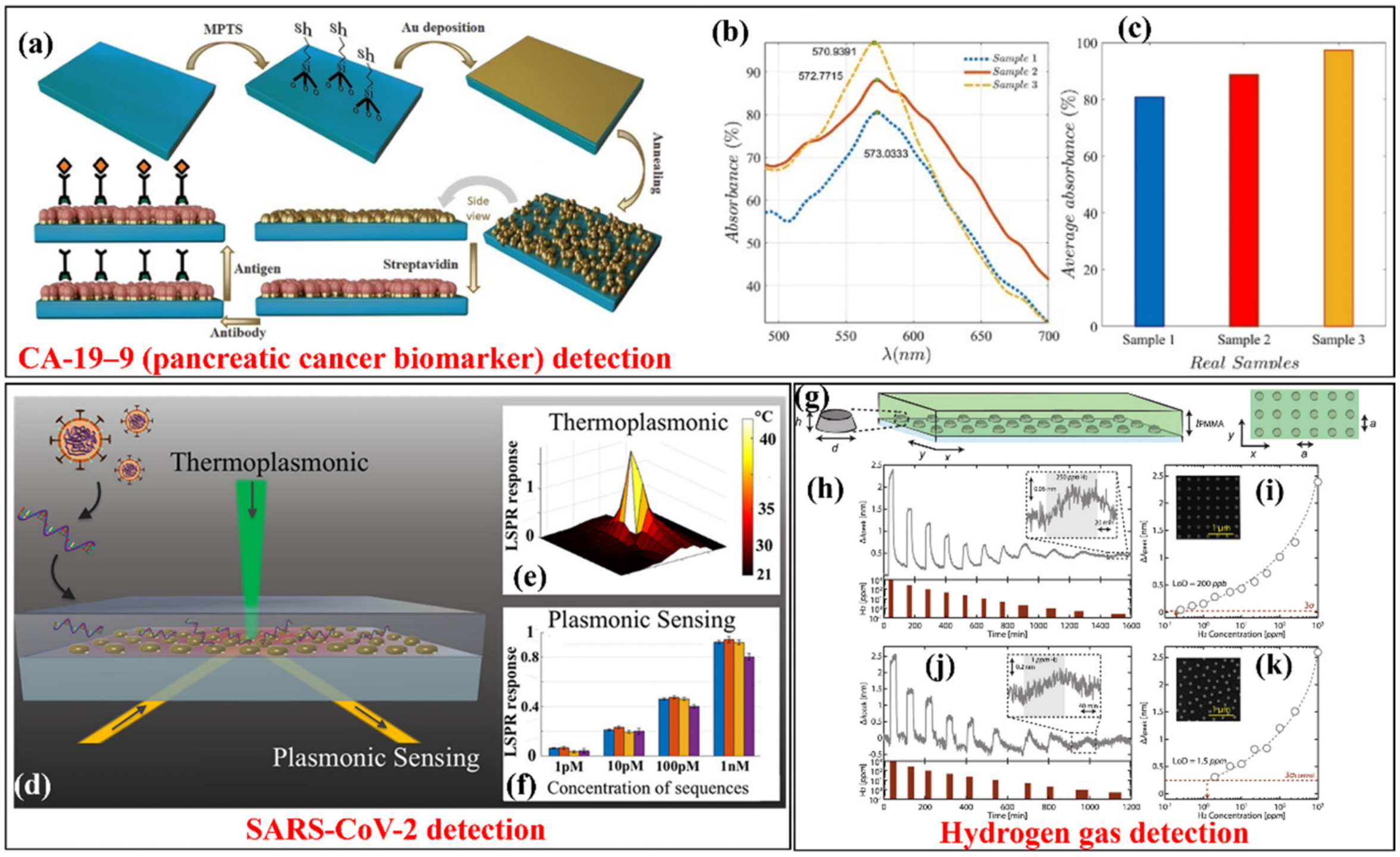

**FIG. 7.** Cutting-edge experimental applications exploring the use of localized surface plasmon resonance (a) Stepwise preparation of a gold nanoparticle array on a glass substrate. (b) and (c) Detection of Carbohydrate antigen 19–9 (CA-19–9) in the three different samples.[102] (a)–(c) reprinted with permission from B. Khalizadeh *et al.*, Microchem. J., **190**, 108698 (2023). Copyright 2023 Elsevier.[102] (d) Schematic of a plasmonic photothermal enhanced LSPR sensor. (e) Thermal map distribution around the photothermal heat source. (f) Quantification of viral oligos with dual-functional LSPR biosensors [(d)–(f) were reproduced from Ref. 103] (g) Schematic of palladium nanoparticle periodic array. (h)–(k) Part-per-billion level gas demonstration of the sensor.[104] (g)–(k) reprinted with permission from F. A. A. Nugroho *et al.*, Nat. Commun. **13**(1), 5737 (2022). Copyright 2022 Authors, licensed under a Creative Commons Attribution (CC BY) license.[104]

The various sensing methods discussed above exhibit distinct advantages and drawbacks. SPR/LSPR sensing stands out for its label-free nature and straightforward setup. However, they are also susceptible to environmental disturbances. SPR imaging, with its label-free approach and capacity for multiple channels, is advantageous; however, it has channel crosstalk issues that limit its resolution. On the other hand, SERS presents itself as a high-sensitivity option with selectivity for different biomarker species, albeit at the expense of requiring costly equipment. The integration of multiplex capabilities is beneficial for reducing the bioreaction duration and enhancing molecular binding by enabling the simultaneous detection of multiple analytes using the same excitation laser wavelength; however, it requires a complex system setup. Each method offers unique benefits while necessitating careful consideration of their respective limitations. Considering the uniqueness and effectiveness of plasmonic-based biosensors in recent years as promising alternatives to conventional sensors, we listed a few reports in Table III.

## IV. PROGRESSING TOWARD ON-CHIP SPECTRAL SENSING

### A. On-chip absorption-based sensing

On-chip absorption spectroscopy, a form of direct absorption spectroscopy (DAS), is an emerging technique that utilizes integrated photonic devices to conduct spectroscopic analyses directly on minute samples of materials within a microchip. This technology holds immense promise for various applications ranging from chemical sensing and environmental monitoring to biomedical diagnostics. By miniaturizing the entire spectroscopy setup onto a chip, on-chip absorption spectroscopy offers significant portability, cost-effectiveness, and scalability to large-scale deployment advantages, which are highly beneficial for field applications and point-of-care diagnostics, where real-time on-site analysis is crucial.

Similar to the bulk spectroscopy method, on-chip absorption spectroscopy works on the Beer-Lambert law,[127] as given below:

$$I = I_0 \exp^{-\alpha\gamma L}. \tag{2}$$

In the given context, $I_0$ represents the initial intensity of the incident light, $\alpha$ denotes the absorption coefficient of the medium, $L$ represents the interaction length, and $\gamma$ is the medium-specific absorption factor, characterizing the impact of enhanced light-matter interaction due to dispersion. In traditional free-space systems, $\gamma$ is equal to 1, necessitating a significant value of $L$ to achieve sufficient sensitivity in measuring the intensity ratio $I/I_0$. However, in the case of on-chip systems, where $L$ is significantly reduced, a high value of $\gamma$ is important to compensate for this reduction in length. The on-chip system can tailor $\gamma$ as per perturbation theory,[64,128] as given below:

$$\gamma = f\frac{c/n_{cl}}{v_g}, \tag{3}$$

where $c$ denotes the velocity of light in free space, $v_g$ represents the group velocity of the guided mode, $n_{cl}$ is the refractive index of the cladding (i.e., the sensing medium, usually air or water), and $f$ signifies the filling factor, indicating the relative fraction of the optical field present in the sensing medium. Consequently, the sensitivity of the engineered sensor could be amplified by adjusting the group velocity and filling factor.

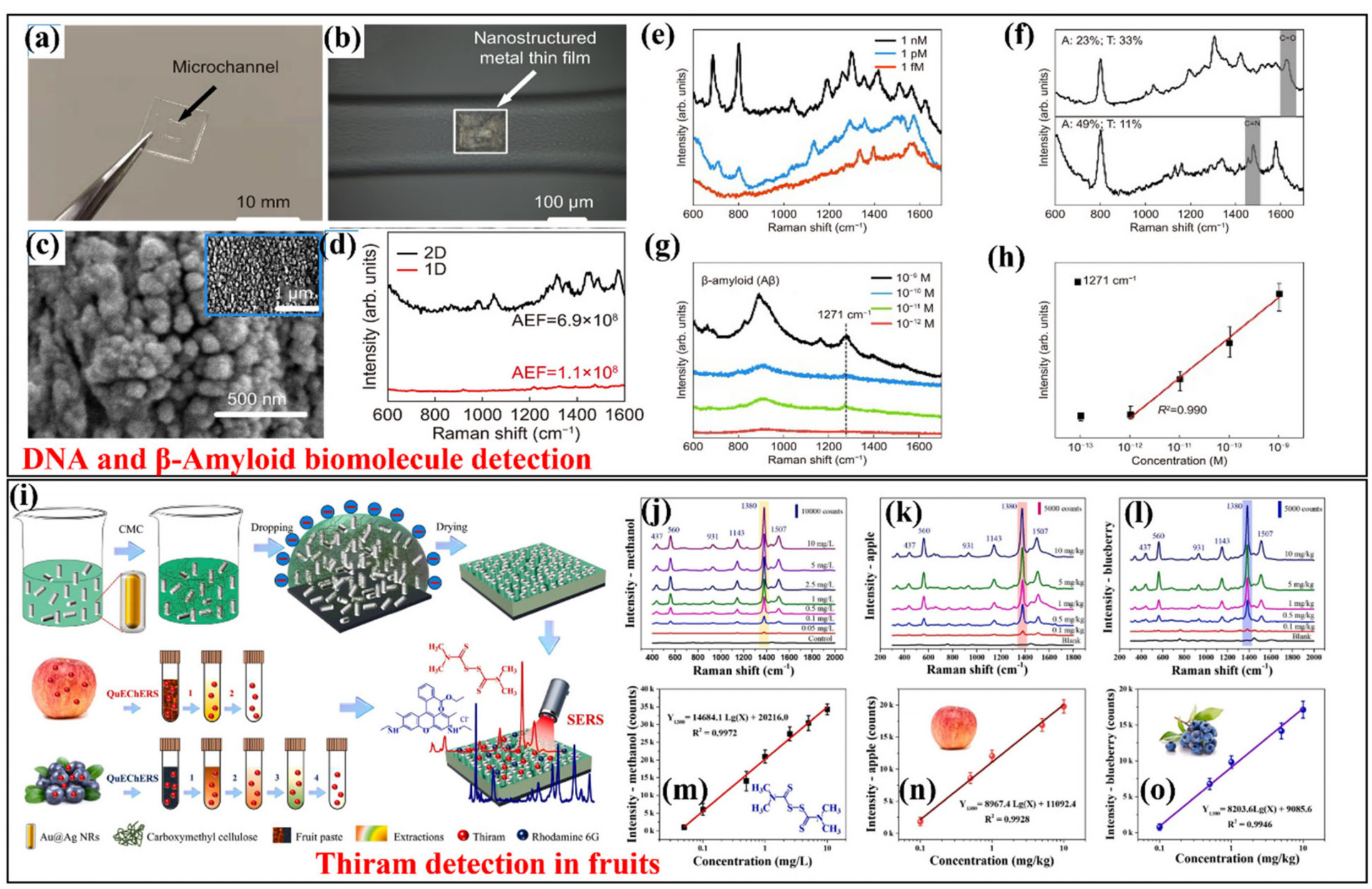

**FIG. 8.** Microfluidic SERS chip created using hybrid femtosecond laser processing. (b) SERS substrate within the glass microchannel. (c) Nanodots formed by secondary laser scanning. (d) Raman spectra of $10^{-9}$ M Rhodamine 6 G (R6G) on 2D (black) and 1D (red) nanostructured SERS substrates. (e) Li-SERS Raman spectra of the DNA sequences at varying concentrations. (f) Raman spectra for 10 nM DNA sequences with different base ratios, highlighting peaks at 1640 $cm^{-1}$ (C = O) and 1474 $cm^{-1}$ (C = N). (g) Raman spectra of A$\beta$ (29–40) at different concentrations using LI-SERS, (h) demonstrating a linear relationship of Raman intensity at 1271 $cm^{-1}$ with concentration (red line).[119] (a–h) Reprinted with permission from S. Bai *et al.*, Opto-Electronic Adv., **5**(10), 210121–1 (2022). Copyright 2022 Authors, licensed under a Creative Commons Attribution (CC BY) license.[119] (i)-(o) SERS spectra of thiram at different concentrations in methanol (j), apple (k), and blueberry (l) using the Au/Ag NRs-CMC chip; (m-o) Correlation between SERS intensities of thiram at 1380 $cm^{-1}$ and the logarithm of thiram concentration in methanol (m), apple (n), and blueberry (o), respectively.[120] (i-o) Reprinted with permission from B. Hu *et al.*, Food Chemistry, **412**, 135332 (2023). Copyright 2023 Elsevier.[120]

Initially, researchers favored near-infrared (NIR) spectroscopy because of its advantages, such as reduced absorption by water and higher tissue penetration with readily accessible sources and detectors, making it suitable for various biomedical and analytical applications.[128–131] The multipass cell configuration is a common element that has been extensively utilized to enhance the interaction of light with analytes in absorption-based spectroscopy methods. Guo *et al.*[132] proposed a compact portable laser sensor using a bulk-optics multipass cell, selectively detecting trace ammonia ($NH_3$) at 1.51 $\mu$m and attaining a minimum detection limit of 0.16 ppm through Allan deviation analysis. However, achieving high sensitivity without employing a bulky and alignment-sensitive multipass gas cell to enhance the interaction length is challenging when focusing on portable sensing systems. Fully on-chip absorption spectroscopy was demonstrated by Tombez *et al.* at a wavelength of 1.65 $\mu$m for methane sensing using an evanescent coupling mechanism with a curly-designed optical waveguide. Their results showed detection limits below 100 ppm by volume.[131] However, hundreds of ppm-level detections are insufficient for several practical applications. There is certainly a margin for scientific innovation in addressing this need. In this context, an engineered photonic waveguide holds great promise for substantially enhancing the interaction between light and analyte (i.e., the enhancement factor $\gamma$) through improved light confinement and/or dispersion engineering while maintaining a relatively small on-chip footprint. Based on a dispersion-engineering approch, our group has demonstrated a 300 $\mu$m-long Si photonic crystal slot-defect waveguide NIR absorption sensor at around $\lambda = 1.69\,\mu$m for the detection of xylene in water and reported concentrations down to 0.1 ppm (0.086 mg/l).[130]

**TABLE III.** State-of-the-art derived from diverse plasmonic-based biosensors.

| Sensing Method | Targeted Biomarker | Limit of Detection | Application |
|---|---|---|---|
| SPR | ErbB2 | 3.8 ng/ml | Breast Cancer Detection[121] |
| SPR | Tropomyosin | 1 $\mu$g/ml | Seafood Allergens Detection[122] |
| LSPR | Glypican-1 | 400 particles/ml | Pancreatic Cancer Detection[123] |
| LSPR | SARS-CoV-2 | 319 Copies/ml | COVID-19 Detection[124] |
| SERS | Shiga Toxin (Stx2) | 0.49 aM (Enhancement Factor 1010) | Foodborne Pathogen Detection[125] |
| SERS | Bilirubin | $10^{-9}$ M | Diagnosis of Jaundice[126] |

To follow further advancement in the absorption spectroscopy technique, the mid-infrared (MIR) region (wavelengths from 2 to 20 $\mu$m) has proven to be a turning factor for the on-chip absorption sensor. The MIR offers enhanced sensitivity because most molecules (specifically, those with polar bonds) exhibit their fundamental vibrational modes in this region, yielding absorption signatures 10–1000 times those in the NIR region.[133] Moreover, absorption signatures above $\sim$7 $\mu$m are incredibly detailed and unique, affording enhanced molecular specificity (leading to this region being called the "fingerprint region").

For on-chip spectroscopic sensors, exploiting silicon's optical transparency from 1.1 to 8 $\mu$m, most photonic sensing waveguides are constructed in silicon-on-insulator (SOI). These waveguides are frequently preferred owing to their compatibility with complementary metal-oxide-semiconductor (CMOS) technology. Additionally, they provide high-index contrast guidance, facilitating sub-micron-sized waveguides and thereby small footprints. However, the transparency of the SOI-based platform is restricted to up to 3.7 $\mu$m due to the presence of oxide cladding. An alternative approach involves the use of freestanding silicon membranes, enabling the utilization of the entire low-loss silicon spectrum.[134,135] In one such approach, Briano *et al.* illustrated the detection of $CO_2$ gas at a 4.24 $\mu$m wavelength, with a $CO_2$ sensitivity approaching 44% of that achieved in free-space sensing [Figs. 9(a)–9(d)]. However, the fabrication of such suspended structures is challenging, resulting in concerns regarding their durability. Our group engaged in this effort has conceptualized an exclusive vertical photonic crystal waveguide entirely made of silicon. By leveraging the photonic bandgap, slow light within the air-core defect enables simultaneous multiplex sensing in a single shot using affordable mid-IR LEDs.[136–138] Researchers are also exploring other promising material platforms like $Si_3N_4$ and silicon on sapphire (SoS)[139,140] to avail high contrast core-clad choice with an extended low-loss window in the Mid-IR regime. Several advanced materials have also been utilized for this purpose.[141,142] For example, Vlk *et al.* presented an integrated waveguide sensor for mid-infrared spectroscopy with a notable 107% evanescent field confinement in the air. Experimental validation at 2.566 $\mu$m demonstrated a remarkable detection limit of 7 ppm for acetylene, utilizing a 2 cm long $Ta_2O_5$ waveguide[141] [Figs. 9(e)–9(h)]. Various works have optimized structures for improved optical confinement in the cladding/sensing region (i.e., increased filling factor $f$).[134,143–146]

In another absorption spectroscopy approach, Zhou *et al.* employed a microring resonator array for multiwavelength mid-IR sensing to guide light from a broad source into distinct wavelength channels. Using small microrings on silicon-rich silicon nitride films, the system achieves a sizable free spectral range of 100 nm across the four channels and selectively detects hexane and ethanol vapor pulses in their mid-IR absorption bands [Figs. 9(i)–9(k)].[147]

The seamless integration of light sources with photodetectors and low-loss optical sensing waveguides is valuable, offering an on-chip spectroscopic sensor that is insensitive to alignment variations and resilient to vibrations. The selection of the materials plays a crucial role in this integration. The most efficient operation of quantum cascade lasers (QCLs) has been demonstrated on an InP platform. The well-matched lattice constant of $In_{0.53}Ga_{0.47}As$ with InP materials exhibits optically transparent characteristics across a broad spectrum range ($\lambda = 3$–$15\,\mu$m).[150] Notably, the InGaAs/InP platform is the optimal choice, allowing for epitaxial growth of QCL/QCDs on the same wafer without relying on precarious wafer/chip bonding processes. Numerous attempts have been made to realize an absorption sensor through monolithic integration on this platform, as evident in previous studies.[64,149,151,152] Utilizing the InGaAs/InP platform, our group has experimentally detected 5 ppm ammonia using a 1 mm suspended holey photonic crystal waveguide and a 3 mm suspended subwavelength grating cladding waveguide. Additionally, we estimated a minimum detectable gas concentration of 84 ppb based on the Beer–Lambert infrared absorption law[148] [Figs. 9(l)–9(o)]. Recently, Hinkov *et al.*[149] introduced a compact chip-scale sensor utilizing quantum cascade technology for real-time monitoring of molecular dynamics in liquid solutions. Their fingertip-sized device integrates emitter, sensing, and detector functions on a single chip. They demonstrated excellent performance with high absorbance (55 times more than a bulk optics system) and a wide concentration coverage (75 $\mu$g ml$^{-1}$ to 0.092 g ml$^{-1}$) [Figs. 9(p)–9(u)]. Nevertheless, it is important to highlight that a drawback of employing such a monolithic approach is that, if any individual component malfunctions or if the sensing waveguide becomes contaminated, the entire chip must be discarded, particularly in liquid-type analyte sensing scenarios. This aspect is of significant importance when considering cost estimation.

### B. On-chip Fourier transform spectrometers

Fourier transform spectroscopy (FTS) is a versatile technique capable of analyzing a broad range of materials, including solids, liquids, and gases. The principle of FTS involves generating a spectrum by applying a Fourier Transform to a time-domain interferogram produced by an interferometer, such as a Michelson interferometer (MI) or lamellar grating interferometer. Compared to dispersive optical spectrometers, FTS offers advantages such as wavelength multiplexing, high optical throughput, high resolution, and a high signal-to-noise ratio (SNR). However, conventional free-space optics FTSs are bulky and expensive, limiting their applicability in various fields, such as remote sensing, airspace exploration, environmental monitoring, meteorological monitoring, biomedical science, and nanotechnology. In effect, there is a growing demand for miniaturized on-chip FTS devices, driven by the need for high sensitivity, cost-effectiveness, noninvasiveness, and real-time detection capabilities, particularly for sensing applications. Integrating FTS components onto microchips presents challenges, especially in implementing MI configurations with moving parts, owing to limited component travel ranges. Nonetheless, the miniaturization of FTS to on-chip platforms promises a transformative shift in analytical capabilities, enabling the development of compact, portable devices that enhance accessibility and reduce costs. Moreover, integrating FTS components into chips for sensing applications paves the way for integration with other on-chip technologies, such as sensing elements, as described in this review. This advancement holds promise for the development of highly integrated and efficient analytical sensing systems with diverse applications. This section explores the recent advances, challenges, and applications of on-chip FTS devices, emphasizing the development of efficient and adaptable analytical tools.

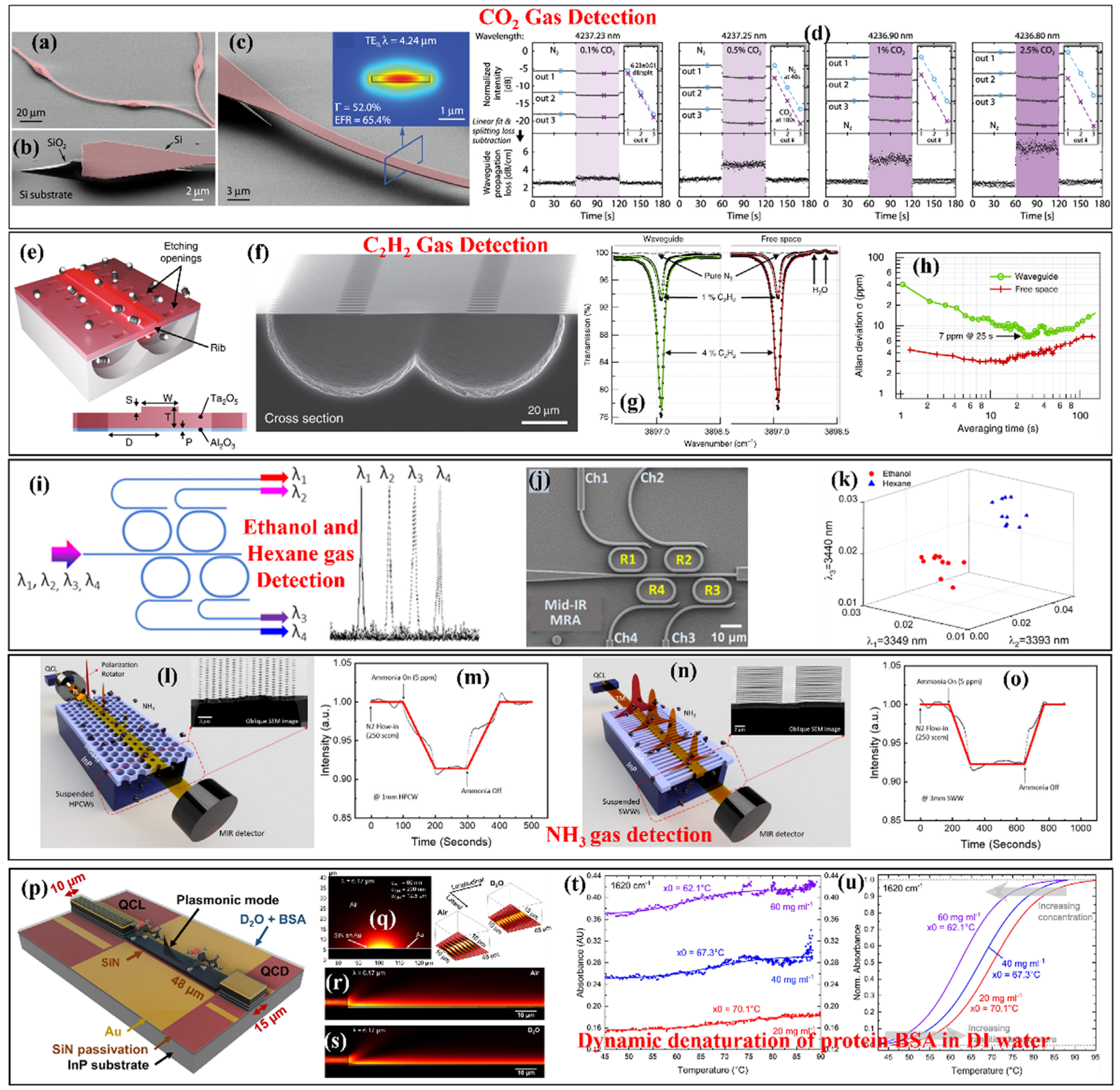

**FIG. 9.** Recent advancements in on-chip absorption sensors. (a)–(c) SEM image of suspended silicon waveguide operating at $\lambda = 4.24\,\mu$m, inset shows the simulated TE polarized mode profile. (d) gas absorption characteristics with 0.1%, 0.4%, 1%, and 2.5% $CO_2$ concentrations (top panel) and respective propagation loss (bottom panel).[134] (a)–(d) reprinted with permission from F. Ottonello-Briano *et al.*, Opt. Lett. **45**(1), 109–112 (2019). Copyright 2019. The Optical Society.[134] (e) Shallow rib tantalum pentoxide membrane core ($Ta_2O_5$), bottom-passivated with $Al_2O_3$ designed to support TM optical mode. (f) SEM of fabricated device. (g) Measured spectra using a 2 cm long waveguide (left) and in free space equivalent path length (right) at various $C_2H_2$ concentrations in $N_2$. (h) Allan deviation plot corresponding to the concentration data time series for 1% $C_2H_2$.[141] (e)–(h) reprinted with permission from M. Vlk *et al.*, Light **10**(1), 26 (2021). Copyright 2021 Authors, licensed under a Creative Commons Attribution (CC BY) license.[141] (i) Depiction of four-channel MRR operating at distinct wavelengths. (j) SEM image. (k) 3D absorption plot of hexane (blue) and ethanol (red).[147] (i)–(k) reprinted with permission from J. Zhou *et al.*, Anal. Chem. **94**(31), 11008–11015 (2022). Copyright 2022 American Chemical Society.[147] (l) diagram depicts a suspended hollow photonic crystal waveguide and cross section image in the inset. (m) Ammonia detection at 5 ppm with TE polarized propagating mode. (n) illustration of suspended sub-wavelength grating-based waveguide and SEM image (inset). (o) Ammonia detection at 5 ppm with TM polarized propagating mode.[148] (p) Pictorial view of monolithic integration of QCL/QCD with the plasmonic waveguide. (q)–(s) Simulated plasmonic mode in the waveguide. (t) Denaturation measurement of BSA at different concentrations. (u) Evaluation of Boltzmann-fit curves illustrates the temperature- and concentration-dependent characteristics of the sigmoidal-shaped absorbance curve.[149] (p)–(u) Reprinted with permission from B. Hinkov *et al.*, Nat. Commun. **13**(1), 4753 (2022). Copyright 2022 Authors, licensed under a Creative Commons Attribution (CC BY) license.[149]

The use of microelectromechanical systems (MEMS) technology has significantly advanced state-of-the-art on-chip FTS devices.[153] MEMS mirrors have replaced the bulky scanning modules of traditional Michelson Interferometer (MI)-based FTS systems, leading to the emergence of various MEMS-based FTS devices with different actuation mechanisms. With a special emphasis on the integration of MEMS technology aimed at miniaturization, portability, and improved spectral resolution, researchers have explored various MEMS-based micromirror designs, including comb-driven actuator techniques,[154,155] MEMS micromirrors with out-of-plane electrostatic capabilities,[154,156] and lamellar grating actuators,[154,157,158] demonstrating high performance and portability. In FTS, the two main performance parameters of interest are the broad spectral range and fine spectral resolution. To address these challenges in chip-scale optical spectrometers, Fathy *et al.*[159] utilized multiple parallel interferometers to achieve a single MEMS actuator-based infrared spectrometer on a silicon chip. They carried out effective monitoring and successfully demonstrated distinct methane absorption bands in greenhouse gas analysis. However, the complexity and challenges of fabrication increase when multiple interferometers are involved [Figs. 10(a)–10(d)]. Qiao *et al.* present a method involving a MEMS-based on-chip computational spectrometer designed for mid-infrared sensing in the range of 3.7–4.05 $\mu$m. By leveraging time-domain modulation of reconfigurable waveguide couplers, their approach presents advantages such as low power consumption, single-pixel detection, and spectral multiplexing capabilities. This study shows the analysis of a broad absorption spectrum of nitrous oxide gas, achieving a spectral resolution of 3 nm [Figs. 10(e)–10(k)].[160]

MEMS-based on-chip FTS systems offer advantages but face challenges such as a limited mirror range, susceptibility to environmental conditions, complex fabrication, durability concerns, scaling issues, alignment requirements, and optical tradeoffs. These factors affect the spectral resolution, performance, and long-term reliability. As an alternative approach, researchers have explored on-chip static FTS devices that involve electro-optic/thermo-optic-based tuning of the optical signal by changing the waveguide refractive index. These approaches avoid moving parts, such as stationary wave integrated FTS (SWIFTS) and spatial heterodyne FTS (SHFTS). The SWIFTS approach involves creating a fixed wave pattern through the interference of either opposing or parallel-propagating waves.[161–163] Coarer *et al.*[162] utilized SWIFTS for the first time to develop a highly compact one-dimensional integrated spectrometer, demonstrating a resolution of 4 nm and a bandwidth of 96 nm using a single stationary MZI. However, this type of FTS device has the limitation of narrow bandwidth owing to the limited optical path delay (OPD) and pitch distance of the detectors for detecting standing waves. In contrast, the SHFTS achieves high resolution through an array of stationary MZIs with a linearly increasing optical path delay or difference. A fixed propagation phase delay is achieved through a combination of multiple elements that provide stepwise phase shifts and/or by thermally or electrooptically modifying the effective refractive index of the waveguide to achieve the desired phase shift. Our group demonstrated several on-chip FTS devices on a variety of material platforms, including SoS, Si, and $SiN_x$. In one such endeavor, Heidari *et al.*[164] demonstrated an on-chip FTS on silicon-on-sapphire featuring 12 MZIs with a linearly increasing OPD. The strip waveguides exhibited a propagation loss of 5.2 dB/cm, and the optical spectrum retrieval from an interband cascade laser at 3.3 $\mu m$ demonstrated a resolution better than 10 $cm^{-1}$.

The spectral resolution and bandwidth of the SHFTS depend on the number of MZIs and maximum OPDs. Striking a balance is imperative to accommodate the constraints of limited chip scale and detection conditions. To achieve unprecedented resolution, various designs of static on-chip FTS devices based on unbalanced MZI arrays have been investigated.[165] For instance, a standard SHFTS device demonstrated earlier featured a 24 MZI array, offering a 5 nm resolution and a 60 nm bandwidth centered at $\lambda_0 = 900$ nm on the $Si_3N_4$ platform.[164] Zheng *et al.*[166] demonstrated a microring-resonator-enhanced FTS device. This innovative device incorporates a tunable MZI linked with a thermally tunable MRR to significantly improve resolution. Notably, the MRR elevates the resolution to an impressive value (0.47 nm, surpassing the Rayleigh criterion for tunable MZI-based FTS [Figs. 11(a)–11(d)]. In addition, researchers have explored large-scale tuning of network structures. Yao *et al.*[167] utilized a mesh of thermally tuned MZIs, achieving a resolution of 20 pm with a bandwidth of 115 nm. In another innovative approach to overcome the aforementioned trade-off between bandwidth and resolution, Min *et al.*[168] reported broadband optical spectrometers utilizing a subwavelength grating coupler integrated with an SHFTS device based on an unbalanced MZI array. To overcome this trade-off, the proposed approach uses bandpass sampling to reconstruct narrow-band channels and achieves a broad-spectrum coverage of 400 nm (650–1050 nm) with 2–5 nm resolution [Figs. 11(e)–11(h)]. Li *et al.*[169] address the bandwidth-resolution trade-off effectively through an innovative computational spectrometer. Unlike conventional resonator-based approaches, they employ a unique type of broadband filter known as a multipoint self-coupled waveguide (MPSCW). By leveraging the low-linear dependency among 64 filters, they successfully attain a broad bandwidth of 100 nm while maintaining a fine resolution of 0.1 nm [Figs. 11(i)–11(k)].

Continuing with advancements in digital spectrometry, Kita *et al.* demonstrated a high-resolution and scalable on-chip digital FTS device and proposed digital signal-processing techniques and optical switches to dynamically select discrete OPDs, enabling precise spectral analysis with exponential scaling of both resolution and channel count.[170] In another approach, Montesinos-Ballester *et al.*[161] demonstrated an alternative on-chip SHFTS approach that utilized spatial heterodyning and optical path tuning via the thermo-optic effect. This approach effectively overcomes the bandwidth-resolution trade-off encountered in conventional devices. This method involved an array of asymmetric MZIs with both incremental and common thermally tuned OPDs. The output interferogram captures data for various temperatures, enabling fine resolution and broad operation without the need for large interferometers. A mid-IR spectrometer utilizing Ge-rich SiGe waveguides achieved a fourfold increase in bandwidth ($603\,\text{cm}^{1}$), while also attaining fine resolution ($<15\,\text{cm}^{1}$). Podmore *et al.*[171] demonstrated a slow-light-induced OPD in an MZI-based FTS device on a photonic chip using a spatial heterodyne configuration and a compressive-sensing technique. The MZI arms were designed to have a group index difference along with the required phase delay by controlling the length of the subwavelength region in the delayed arm. Compressive sensing then exploits the spatial interferogram sparsity or compressibility (downsampling) of the spatial interferogram, enabling accurate reconstruction of the input spectra with full resolution and bandwidth using only 14 MZIs [Figs. 10(i)–10(j)]. To perform a sensing analysis utilizing an FTS device, Yoo *et al.*[172,173] integrated an MRR and an on-chip FTS device to eliminate the need for an external optical spectrum analyzer. The output resonance peaks reconstructed through the integrated SHFTS device provided a spectral resolution of approximately 3.1 nm, a bandwidth of approximately 50 nm, and a remarkable limit of detection of 0.042 RIU. The integration of these components on a single SOI wafer streamlines the sensing process while enhancing the overall performance and sensitivity of the biosensor system.

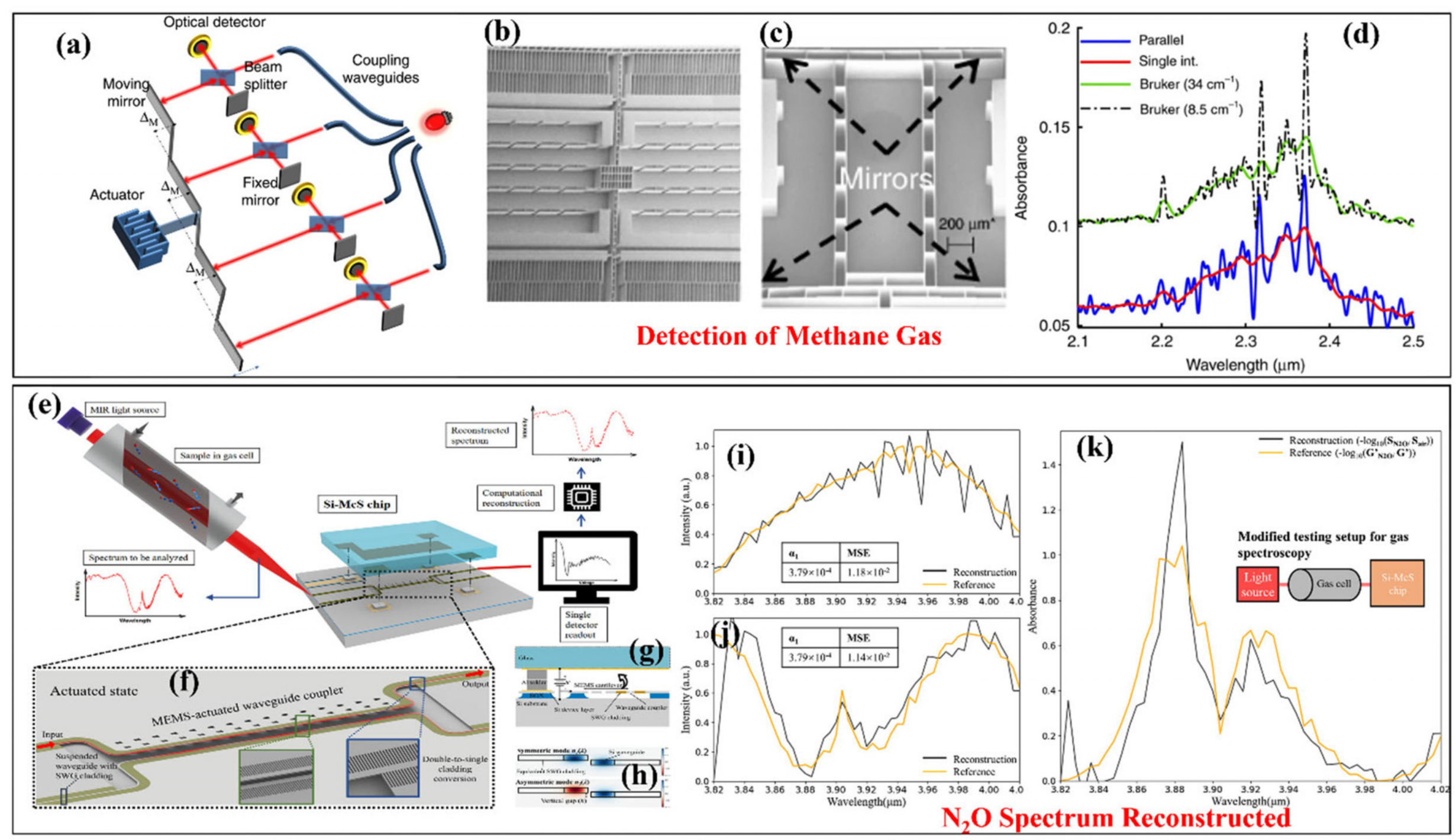

**FIG. 10.** MEMS-based on-chip FTSs. (a) Illustration of parallel interferometers sharing a common actuator featuring spatially shifted moving mirrors to acquire shifted interferograms. (b) SEM image of the actuator. (c) SEM image depicting the moving mirror arm. (d) Comparative analysis of methane absorbance curves measured by parallel interferometers (blue) and a single interferometer (red), in contrast to the absorbance from a benchtop spectrometer for corresponding resolution.[159] (a)–(d) Reprinted with permission from A. Fathy *et al.*, Micro. & Nanoengg., **6**(1), 10 (2020). Copyright 2020 Springer Nature.[159] (e) MIR spectroscopy system employing the silicon MEMS-enabled computational spectrometer chip (Si-McS): the MIR light traverses the gas cell, and the absorbed light is analyzed on the Si-McS chip through computational reconstruction based on the measured interferogram. (f) Close-up of the spectrometer with the actuation element, showing upward movement of the MEMS cantilever. (g) Cross section view illustrating applied bias and relative motion during operation. (h) Field distribution of propagating mode profiles of the coupler. (i) and (j) Reconstruction of the spectra of air and $N_2O$ gas, respectively. (k) Calculated absorbance reveals spectral features indicative of detected $N_2O$ gas.[160] (e)–(k) reprinted with permission from Q. Qiao *et al.*, ACS Photonics **9**(7), 2367–2377 (2022). Copyright 2022 American Chemical Society.[160]

### C. Wavelength modulation spectroscopy

Wavelength modulation spectroscopy (WMS) has emerged as a powerful and versatile spectroscopic analysis technique that offers notable advancements over conventional methods, such as direct absorption spectroscopy. This section provides an overview of the WMS principle, advantages, and applications, with a particular focus on the recent developments in higher-harmonic detection and its integration into photonic on-chip systems. The WMS technique enhances the detection signal-to-noise ratio by modulating the laser frequency (or wavelength) at a specific frequency and amplitude, enabling precise spectral analysis and selective detection of the molecular species of interest. Sinusoidal modulation improves the SNR of measurements by reducing low-frequency noise (1/f) contributions.[174–176] The modulated laser frequency ($\nu(t)$) can be mathematically formulated as shown below:[177,178]

$$\nu(t) = \nu_c + \nu_m \cos(2\pi f_m t), \tag{4}$$

where $\nu_c$ and $\nu_\mathrm{m}$ are the laser center frequency (typically in tens to hundreds of THz) and modulation depth (typically in a few THz), respectively, and $f_m$ represents the modulating frequency (typically in kHz). The modulated laser beam traverses through the sample, inducing variations and changes in transmitted light intensity at the modulation frequency ($f_m$). Unlike the conventional approach, where the transmittance, $T = I(\nu)/I_0$ is measured, the WMS technique quantifies relative absorption, $\bar{A}$ in terms of the incident intensity ($I_0$) and the intensity of light after transmitting through the absorption medium ($I$), which is defined as:[177,178]

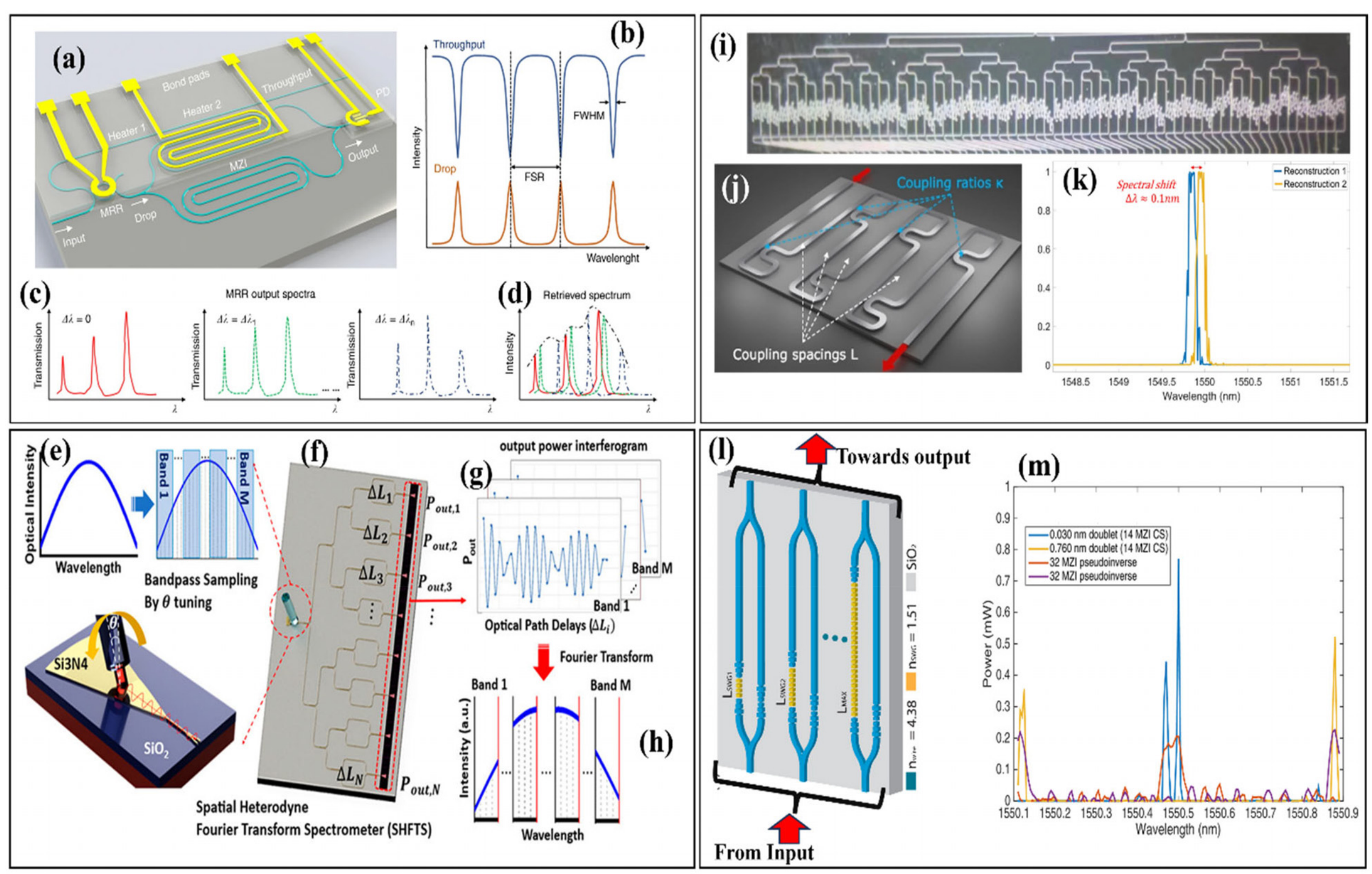

**FIG. 11.** Various on-chip Fourier transform spectroscopy methods are illustrated: (a) Microring resonator-assisted Fourier-transform spectrometer with an MRR and MZI integrated with a PD, (b) MRR transmission spectra, (c) Output filtered spectra from the MRR, (d) Retrieved spectra from the tunable MZI.[166] (a)–(d) Reprinted with permission from N. Zheng *et al.*, Nat. Commun. **10**(1), 2349 (2019). Copyright 2019 Authors, licensed under a Creative Commons Attribution (CC BY) license.[166] (e) SWGC operation as a coupler and tunable-bandpass filter, (f) $Si_3N_4$ SHFTS chip with N-MZIs, (g) Spatial interferograms from MZI array outputs, (h) Reconstructed narrow-band spectra by Fourier transform,[168] (i) Microscopic images of the computational spectrometer, (j) One of the 64 filters, (k) Application of spectral analysis from fiber Bragg grating sensor, (l) Light directed to an array of interferometers, and (m) spectra retrieval using unapodized interferometric measurements from a subset of MZIs.[169] (i)–(m) Reprinted with permission from A. Li *et al.*, PhotoniX, **4**(1), 29 (2023). Copyright 2023 Springer Nature.[169]

$$\bar{A} = \frac{\Delta I}{I_0} = \frac{I_0 - I}{I_0}. \tag{5}$$

The photodetector records these modulated intensity fluctuations, and its output, which contains both amplitude and phase modulation, is analyzed in a phase-sensitive device known as a lock-in amplifier. This process extracts higher-order harmonics ($Nf_{\mathrm{m}}$) of the modulation frequency (where $N = 1, 2, 3$, etc.). For instance, $2f_m$ is the second harmonic and $3f_m$ is the third harmonic, and so forth. Analyzing these higher-harmonic terms within a Taylor or Fourier series contributes to a more accurate representation of the signal to be detected. The inclusion of higher harmonics enables the capture of distinct absorption line profiles, which is particularly beneficial for overlapping and complex absorption line features.

Various theoretical frameworks have been proposed for the WMS technique.[179–183] Initial studies used a Taylor expansion of the absorption profile to derive higher harmonics of the modulated signal, which was effective when the modulation depth was much smaller than the half-width of the absorption profile.[182–184] However, for larger modulation amplitudes, the Taylor approach becomes invalid, leading Wilson to use numerical integration for any amplitude, resolving the first three harmonics for Gaussian- and Lorentzian-broadened lines.[179] Arndt[180] provided an analytical solution for Lorentzian profiles using Fourier analysis, offering expressions for the $1f_m$ and $2f_m$ harmonics. Reid and Labrie[182] performed the first experimental validations with diode lasers, and Wilson's method was later used for Voigt profiles. Progress in higher-order harmonics detection ($1f_m$, $2f_m$, $3f_m$)[185–188] led to applications in atmospheric science, trace-gas analysis, and material characterization.

WMS offers several advantages that contribute to their widespread adoption in various environmental, biomedical, and industrial applications. Its high sensitivity makes it well-suited for accurate measurements of low concentrations in trace gas analysis using semiconductor lasers in the mid-infrared and near-infrared regions.[174–176,181,189–198] The selectivity of the target analyte species is enhanced by focusing on the specific wavelengths corresponding to the absorption features of the sample. The quantitative nature of WMS analysis, correlating signal characteristics with species concentration, further establishes its utility and applicability in many trace gas-sensing applications at high pressure and temperature. Higher temperatures broaden the absorption lines, increasing the signal strength, while higher pressures enhance collisional broadening and improve the spectral resolution. These factors enable a more accurate detection in challenging environments. The versatility of WMS is reflected in its applications across diverse fields. Environmental monitoring,[197–200] industrial process control,[201–205] and gas sensing benefit from the higher sensitivity and selectivity of the WMS technique. This technique has been used extensively in studies related to atmospheric chemistry, air quality monitoring, and remote sensing applications. Its capability to detect trace gases in various matrices underscores its significance in medical applications, such as human breath analyzers for lung cancer detection.[206–208]

Recent advancements in WMS have extended its capabilities to chip-scale implementations. Studies such as those by Pi *et al.*[209,210] explored the effectiveness of WMS in waveguide sensors compared with DAS. Pi *et al.* explore the use of WMS in waveguide sensors, demonstrating significant sensitivity improvements over direct absorption spectroscopy (DAS). This research focuses on optimizing mid-infrared waveguide methane sensors and fabricating ChG-on-$MgF_2$ sensors, achieving a 24× reduction in the limit of detection (LoD) compared with previous DAS-based sensors [Figs. 12(a)–12(d)].[209] Recently, Zhao *et al.*[211] presented a novel on-chip acetylene ($C_2H_2$) sensor using SU8 polymer spiral waveguides. The sensor utilizes a WMS and incorporates a Euler-S bend design, reducing the sensor size by over 50% in comparison to a normal bend structure. Experimental validation showed a LoD as low as hundreds of ppm, highlighting SU8's promise for compact, high-sensitivity on-chip gas sensing. We envision that WMS continues to evolve, driven by advancements in higher harmonic detection and on-chip implementation [Figs. 12(e)–12(h)]. The sensitivity, selectivity, and noise reduction capabilities of WMS establish it as an asset across various sensing applications in scientific and industrial fields. With ongoing research and technological advancements, WMS techniques integrated with on-chip components are expected to enable new applications, addressing complex challenges in gas sensing and spectroscopic measurements.

### D. Photonic frequency comb spectroscopy: Toward on-chip integration

An optical frequency comb (OFC) is a spectrum consisting of discrete and equidistant spectral lines. OFCs play a crucial role in bridging the optical and radio- or microwave-frequency domains through a process known as frequency synthesis. This process involves linking the stability and precision of an optical frequency comb to a well-established and easily measurable radio- or microwave-frequency reference. OFCs have quickly become attractive for various applications including precision frequency measurements, timekeeping, telecommunication, broadband molecular spectroscopy, gas sensing, and quantum information processing. Early methods for generating frequency combs[212,213] were based on intracavity phase modulation. Subsequently, mode-locked lasers with stabilized repetition rates and carrier-envelope phases have become a common approach.[214–216]

A commonly used mechanism generating evenly spaced spectral lines in a mode-locked laser (MLL) is mathematically described by the comb equation $f_n = f_0 + n \cdot f_r$, where $n$ is an integer, $f_r$ is the comb tooth spacing (equal to the mode-locked laser's repetition rate or the modulation frequency), and $f_0$ is the carrier offset frequency, which is less than or equal to $f_r$ ($f_0 \leq f_r$). The comb equation highlights that each individual optical mode, $f_n$, can be defined by only two parameters: the repetition frequency and laser offset frequency. The laser repetition rate, $f_r$, which represents the inverse of the pulse-to-pulse timing $\tau r$, remains constant across the spectrum due to the cavity resonance modes stemming from the mode-locking mechanism, thereby ensuring the equidistant spectral lines of the OFCs.[217] Measuring $f_r$ is straightforward and can be done using a photodiode and an RF spectrum analyzer to measure it directly.

Measuring $f_0$ experimentally is challenging due to its connection to the optical carrier phase. In 1999, a method was proposed to generate a heterodyne beat at $f_0$ through nonlinear self-referencing between the endpoints of the optical comb spectrum.[218,219] The simplest implementation of heterodyning at $f_0$ involves doubling the frequency of light from a low-end comb mode and then interfering it with fundamental light at twice that frequency to determine $f_0$, such that $f_0 = 2 \cdot (n f_r + f_0) - (2n \cdot f_r - f_0)$. Implementing the mathematical concept of an optical frequency comb requires covering an optical octave of bandwidth, which is challenging because the broadest mode-locked lasers are typically smaller than 100 nm.

In the 1990s, the development of highly engineered low-dispersion optical fibers enabled the generation of a coherent white-light continuum using ultrashort laser pulses. This breakthrough, combined with OFCs from Ti:Sapphire lasers, has led to rapid advancements in precision metrology. Within a few years later, various applications were demonstrated,[213,214,220–225] including atomic clocks, optical frequency measurements, attosecond control, and direct molecular spectroscopy, marking a period of intense progress in the OFC research field. Stabilizing frequency combs involves employing various techniques, including phase locking to external references (e.g., GPS-disciplined oscillators)[226] fiber stretching,[227] feedforward and feedback control[228] (Kärtner *et al.*, 2003), and the use of optical cavity resonators.[229] Additionally, carrier-envelope offset (CEO) stabilization using f-2f interferometers[214] and temperature control of critical components contribute to long-term stability in frequency comb systems.[230]

Since the stabilization of OFCs was realized, advancements in MLL systems have led to the evolution of solid-state and fiber-based OFCs. Diode-pumped solid-state and fiber lasers, particularly Er: fiber OFCs, dominated commercial success, providing near-continuous spectroscopic coverage from 400 nm to ~4 $\mu$m. The state-of-the-art MLL sources have transitioned from Ti:Sapphire laser systems to compact, environmentally stable Er: fiber OFCs, culminating in sub-100 fs Er/Yb:glass lasers that are highly compact and portable. Instead of using mode-locked lasers, the nonlinearity in photonic resonators can generate a frequency comb. A key advantage of this approach is that it can operate at high repetition rates (>10 GHz), which unlock more applications, and photonics integration is possible.

Optical microcavities have enabled new research fields to blossom, particularly in frequency metrology. The concept of microcavities originated from confining light to very small volumes for an extended time.[231] The traditional challenge has been to fabricate devices with high-quality factors (Q-factors). Braginsky and Ilchenko observed ultrahigh-quality factors inside glass microspheres, which triggered research in the area of high-Q microcavities.[232] Confining light in a small volume for an extended time involves nonlinear optics as a key process. For example, second- and third-order nonlinear optics effects are the driving forces for supercontinuum and Kerr comb generation in frequency metrology.[233] Vahala reviewed four optical microcavities based on their light-confinement methods and discussed their diverse future applications.[234] One interaction worth noting is the coupling between the optical and mechanical modes, which naturally occurs in optical microcavities. This has long been of broad interest in the microcavity research field.[235–237] Another interaction inside microresonators is the Kerr nonlinearity, which is a well-known underlying process responsible for self-phase modulation and supercontinuum generation. In the context of microresonators, the nonlinear Kerr process allows an optical frequency comb to be generated, providing a path for practical frequency comb metrology and related applications.[238]

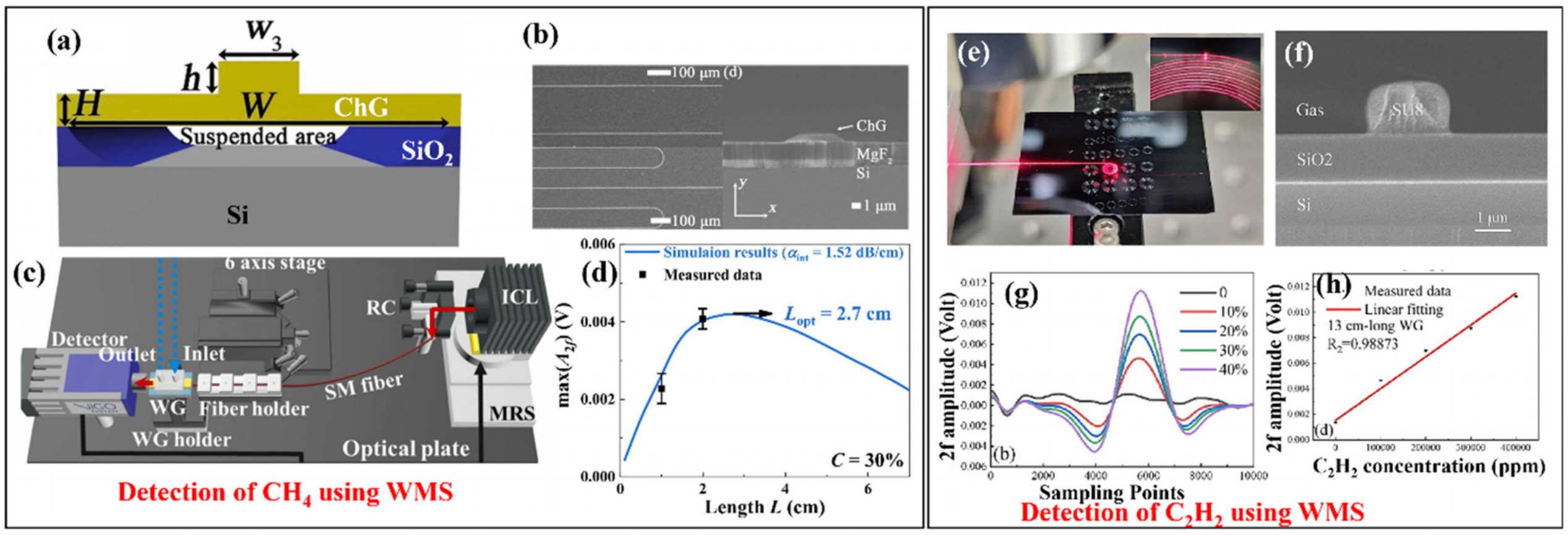

**FIG. 12.** Experimental validation of WMS on a chip scale: (a) and (b) depict the schematic of a suspended optical waveguide made of chalcogenide material on top of a Si platform and the SEM image, respectively. (c) Schematic of measurement system. (d) shows the simulation curve of max($A_2$f) vs L and the measured max($A_2$f) with L = 10 and 20 mm.[209] (a)–(d) Reprinted with permission from M. Pi *et al.*, Sensors Actuators B, **362**, 131782 (2022). Copyright 2022 Elsevier.[209] (e)–(h) Shows the WMS-based $C_2H_2$ sensor, where (e) shows microscopic photos with red light coupled into the spiral waveguide, and (f) shows the cross section electron-microscopic image of the SU8 waveguide on the silicon platform. (g) and (h) show the measured 2f signal under various $C_2H_2$ concentration levels and linear fitting curve for 13 cm-long waveguides.[211] (e)–(h) reprinted with permission from H. Zhao *et al.*, Spectrochimica Acta Part A, **302**, 123020 (2023). Copyright 2023 Elsevier.[211]

Therefore, on-chip optical microresonators have become vehicles for these research fields, including quantum optical phenomena resulting from the coupling of light and mechanical motion, in addition to the optical properties explored in optical microcavities. Alternatively, they can be used to generate frequency combs. The process underlying optical generation involves parametric oscillations and four-wave mixing, where two pump photons are annihilated to generate a single idler photon. According to energy conservation, the excess idler photon needs to be compensated by reducing the signal frequency, thereby creating two equidistant sidebands. This process, opposite to conventional lasers, scales with $1/Q^2$.

Over the past 15 years, notable developments have been made in the field of tiny chip-scale optical frequency comb (OFC) sources. Semiconductor lasers have become highly versatile platforms with a broad wavelength range and the ability to be mass-manufactured at reasonable cost. Examples of these lasers include mode-locked integrated external-cavity surface-emitting lasers (MIXSELs) and QCLs. In particular, MIXSELs exhibit sub-100 fs pulse production and, when optically pumped, over 1 W of optical power. Despite challenges in efficient soliton formation, micro-combs have achieved coherent detection of $f_0$ and expanded to an optical octave at mode spacing <30 GHz. However, thermo-refractive noise remains a hurdle, requiring additional correction mechanisms, such as a blue-detuned auxiliary laser, albeit at the cost of increased complexity and power consumption.

Electro-optic comb generators offer agile mode-spacing tuning and find applications in precision metrology and astronomical spectrograph calibrations. Supercontinuum generation in photonic waveguides overcomes the challenges associated with low pulse energies in certain OFC platforms, thereby presenting a breakthrough in achieving pulse energies below 200 *pJ*. Recent progress in chip-based nonlinear photonics has enabled the development of compact, portable, and fully integrated comb devices with great potential for diverse applications. OFC devices with wide bandwidths are particularly valuable for applications such as spectroscopy and time and frequency metrology, particularly when employing self-referencing via *f–2f* interferometry. Supercontinuum generation (SCG) in optical waveguides and Kerr comb generation (KCG) in microresonators are prominent methods for achieving broad spectral bandwidths.[8,239–245] Supercontinuum generation exploits nonlinear effects in a waveguide to generate a broad comb-like spectrum with evenly spaced frequency components.[245–247]

On the other hand, KCG utilizes the Kerr effect, induced by changes in the refractive index of a material with light intensity, to create equidistant frequency components across a broad spectrum using microresonators. The Kerr nonlinearity within a high-quality (high-Q) microresonator can lead to the formation of soliton pulses, resulting in a comb of equidistant spectral lines. Kippenberg *et al.*[248] reported the first observation of Kerr nonlinearity-induced optical parametric oscillation in a microcavity using geometrical control of its toroidal shape. This study has significant implications for the potential realization of Kerr-based optical frequency comb (OFC) generation on a chip.

Integrated photonic Kerr combs with octave-spanning Optical Frequency Combs (OFCs) have been demonstrated, enabling essential self-reference at $f-2f$.[240,249–251] Savchenkov *et al.*[252] demonstrated low-threshold optical parametric oscillations in a high-Q fluorite whispering gallery mode resonator, stemming from resonantly enhanced four-wave mixing attributed to Kerr nonlinearity.[246] They present the optical parametric oscillations in a high-Q $CaF2$ crystalline resonator, demonstrating the efficient generation of narrow-band optical sidebands for an all-optical frequency reference. Kerr combs in microresonators have been achieved using various materials, including silica ($SiO_2$),[239,248,253–255] silicon nitride (SiN),[227,256–259] Si,[239,260] aluminum nitride (AlN),[239,261] and aluminum-gallium arsenide (AlGaAs).[239,262]

Integrated photonic on-chip frequency combs have attracted interest for building sensing systems for various applications, such as detecting various gases, environmental monitoring, and medical diagnostics. In recent years, advancements in on-chip gas sensors utilizing continuous lasers have been reported.[141,148,263,264] However, simultaneous detection of multiple gases remains challenging. Addressing this, Guan *et al.*[263] have designed an integrated photonic multicomponent gas sensor using a microcavity Kerr frequency comb, enabling multiplexed gas sensing for water vapor ($H_2O$), ammonia ($NH_3$), acetylene ($C_2H_2$), and carbon dioxide ($CO_2$) within the 1–2 $\mu$m range. The sensor features a symmetric double-ring microcavity with a feedback structure for dispersive Kerr soliton generation and a long sensing waveguide to achieve adequate absorption at NIR wavelengths [see Figs. 13(a)–13(b)]. Their study marked the inception of an integrated Kerr frequency comb for on-chip gas sensing, showing remarkable stability and sensing performance evaluations with the promise of ppmv-level sensitivity.

Yu *et al.*[265] presented micro-resonator-based soliton combs utilizing a microfluidic chip-based dual-comb spectroscopy (DCS) method [Figs. 13(c)–13(d)] for real-time probing of linear absorption in liquid acetone within the mid-infrared range (2900–2990 nm). Notably, the experiment achieved a high spectral acquisition rate of 25 kHz, providing insights into acetone droplet dynamics and showing that compact time-resolved spectroscopy systems are applicable in various sensing and imaging domains such as chemistry, biology, and industry. Heterogeneous integration of $Si_3N_4$-based micro-resonators and III-V semiconductor lasers promises cost-effective mass production of frequency comb chips. Progress in self-injection locking and nonlinear effects has driven optical frequency combs from laboratories toward market applications. Integrating key components holds the potential for superminiaturized, fully on-chip direct frequency comb sensing spectral chips, facilitating the practical use of optical frequency combs in various applications.

Table IV provides a concise overview of the capabilities and performance metrics of on-chip spectroscopic methods. This facilitates the comparison and understanding of their potential applications across various domains, from environmental monitoring to biomedical diagnostics.

## V. ENHANCING ON-CHIP BIOSENSOR PERFORMANCE WITH ARTIFICIAL INTELLIGENCE/MACHINE LEARNING ALGORITHMS

Artificial intelligence (AI) and machine learning (ML) play crucial roles in enhancing the functionality and performance of on-chip biosensors. These technologies enable efficient analysis of complex biological data generated by biosensors, aiding the rapid and accurate detection of biomolecules. AI algorithms can recognize patterns, identify trends, and make qualitative or quantitative predictions based on sensors, leading to improved sensitivity and specificity for identifying new biomarkers and known biological targets. ML models can adapt and optimize sensor parameters, thereby enhancing the overall system efficiency. Moreover, AI facilitates real-time monitoring, interpretation of dynamic biological responses, and development of intelligent and autonomous biosensing systems. Integrating AI and ML with on-chip biosensors is a promising approach to advancing diagnostic capabilities and personalized healthcare applications.

In this section, we provide a concise explanation of AI/ML. Our primary goal was to enhance the efficiency of on-chip biosensors by integrating AI/ML to precisely articulate the conclusions derived from it and comprehend potential patterns. Figure 14 shows the relationship between AI, ML, neural networks, and various algorithms along with their primary application scenarios. Supervised learning, which involves training models on labeled datasets, is well-suited for applications where historical data with clear outcomes are available, such as in the quantitative analysis of biomolecules.[269,270] In contrast, unsupervised learning is beneficial for tasks such as clustering, pattern recognition, and association analysis, making it suitable for identifying unknown patterns or anomalies in biosensor data.[271] While some unsupervised learning models, such as clustering algorithms and dimensionality reduction techniques, offer a degree of interpretability, others—particularly more complex models, such as autoencoders—can exhibit a "black box" nature, making it difficult to understand which features they focus on. This black-box issue is not exclusive to unsupervised learning, as many complex supervised models such as DNNs also suffer from limited interpretability. This is particularly significant in clinical settings where identifying these features could offer opportunities for biomarker discovery and diagnostic applications. However, it is essential to invest effort in comprehending the inner workings of the model to ensure scientific validation before considering its clinical utility. Reinforcement learning, which is used for decision-making in dynamic environments, can be applied in biosensor applications that require adaptive responses to changing conditions.[272] The choice of the most appropriate AI/ML approach depends on the specific requirements and characteristics of the on-chip biosensor applications. Criteria for selection include the availability of labeled data, complexity of the analysis, and desired level of interpretability. The steps for adopting different types of AI/ML include defining the problem, collecting and preprocessing data, selecting a suitable algorithm, training the model, and evaluating its performance. Continuous refinement and adaptation of the chosen AI/ML technique are essential to ensure optimal results in the dynamic context of on-chip biosensors.

An artificial neural network (ANN) is a computational model designed based on the structure and operation of the human brain. As shown in Fig. 14, ANNs are a subset of machine learning, can be trained using the three learning modes mentioned above, and are particularly effective in solving complex nonlinear problems. It comprises interconnected nodes arranged in layers, that is, input, hidden, and output nodes. The connections between the nodes possess weights that are tuned during training, allowing the network to acquire knowledge and perform predictions or classifications. ANNs possess the capability to grasp intricate patterns and associations in data, thereby proving to be effective for tasks such as classification, regression, and pattern recognition. They demonstrated proficiency in managing nonlinear relationships and adjusting high-dimensional datasets. Utilizing ANNs, Hamedi *et al.* reported a nanophotonic biosensor to accurately predict electrical signal outputs, demonstrating superior performance ($\mathrm{MSE} = 2.9 \times 10^{-8}$) compared to finite-difference time-domain (FDTD) methods. The ANN model, with inputs including the biosample refractive index, central wavelength, and FWHM, enables efficient pre-optimization for sensitivity and responsivity, revealing optimal conditions for a 735 nm central wavelength and 70 nm FWHM light source.[284]

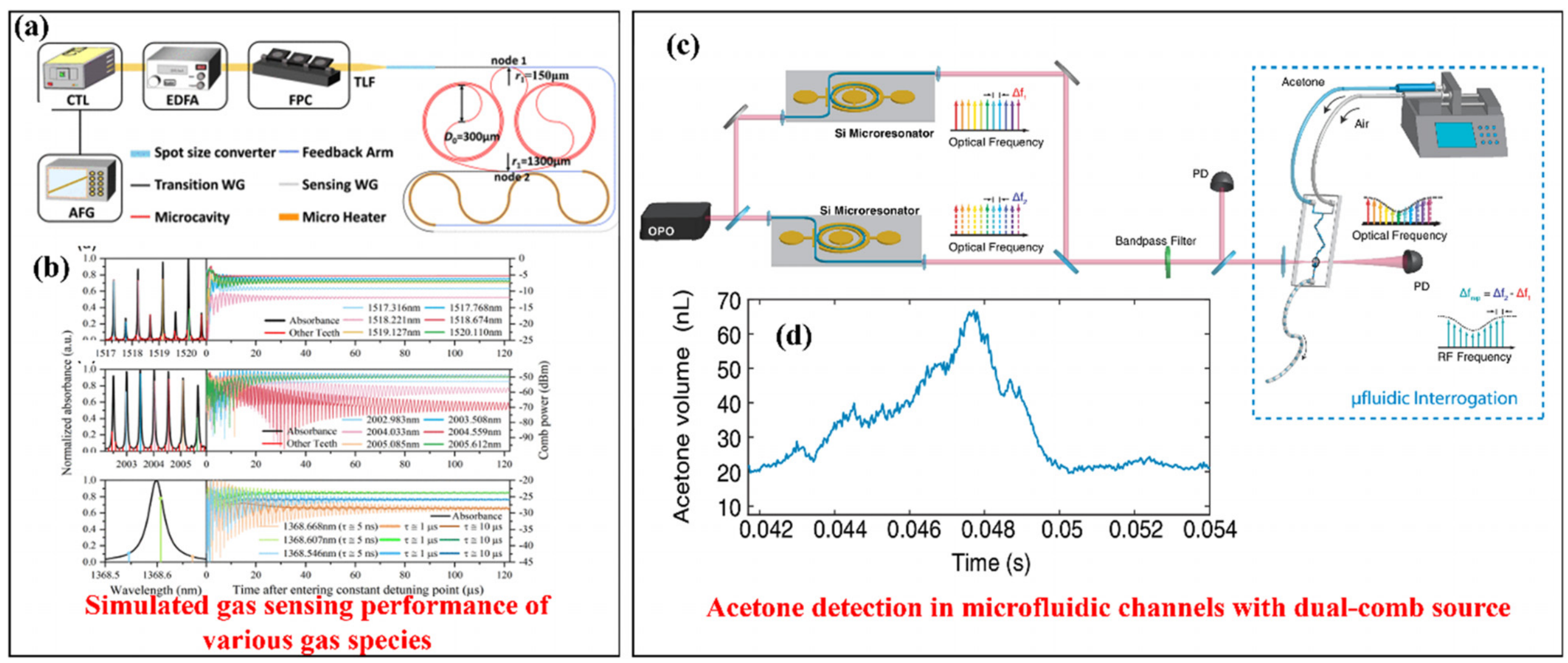

**FIG. 13.** (a) Schematic of the on-chip multi-gas sensor using dissipative Kerr soliton combs. (b) Analysis of the power and thermal stability of the optical frequency comb teeth. Absorption spectra of gases and power fluctuations.[263] (a) and (b) reprinted with permission from Guan *et al.,* Journal of Lightwave Technology, **41**(10), 2308–3224 (2023). Copyright 2023 IEEE.[263] (c) Experimental setup: Continuous-wave OPO is split to pump silicon microresonators for mid-IR frequency comb generation. The outputs were combined, filtered, and split into the reference and sample arms. Acetone droplets were introduced using a microfluidic chip. (d) Calculated acetone volume detected over time with 40 $\mu$s temporal resolution.[265] (c and d) reprinted with permission from M. Yu *et al.*, Opt. Lett. **44**(17), 4259–4262 (2019). Copyright 2019 The Optical Society.[265]

When evaluating different types of AI/ML techniques for on-chip biosensor applications, selection criteria are contingent on various factors. This includes the nature of the data, in which the data type plays a crucial role. Images and complex graphical data might necessitate convolutional neural networks (CNNs).[273–275] For example, a one-dimensional CNN was applied to recognize the absorption spectra of mixtures featuring 64 predefined mixing ratios. The classification accuracy was 98.88%, particularly in the analysis of aqueous mixtures within the mid-infrared (MIR) range using a Si-metamaterial sensing waveguide platform.[276] Alternatively, random forest[277] and XGBoost[278] are widely used models for effectively handling tabular data. Recently, Robison *et al.* developed a 13-plex immunoassay panel to assess cytokine release from peripheral blood mononuclear cells stimulated with Mycobacterium tuberculosis-associated antigens. The study included 65 subjects with diverse TB exposure risks and utilized random forest feature selection to identify cytokine biomarkers[279] detected using silicon photonic microring sensor arrays.

Additionally, the size of the dataset is pivotal because deep learning models often require large datasets for training, whereas simpler models may suffice only for smaller datasets.[280] Feature selection is another critical consideration for identifying the pertinent variables for model training. Furthermore, pre-processing steps, such as data cleaning, denoising, and transformation, must align with the specific characteristics of the data. Common preprocessing techniques include derivatives, denoising, and Fourier transforms. System-specific preprocessing methods include data compression, baseline drift elimination, normalization, transformations, and similar techniques. The adoption of preprocessing methods profoundly influences the overall performance of ML models. In the context of Raman spectroscopy, each spectrum necessitates Savitzky–Golay smoothing, background subtraction, and normalized min-max scaling.[281] This standardized pre-processing ensures that the data are appropriately refined for optimal performance within the ML models. Post-processing steps applied after obtaining predictions may be essential for refining or interpreting results.

**TABLE IV.** State-of-the-art derived from diverse on-chip spectroscopy techniques.

| Spectroscopy type | Wavelength(s) | Sensitivity | LOD/resolution | Analyte detected/targeted |
|---|---|---|---|---|
| DAS | 6.19 $\mu m$ | 0.075–92 mg/ml | ··· | BSA[149] |
| DAS | 3.291, 4.319, and 7.625 $\mu m$ | ··· | 5.9 ppm | $CH_4$ and $CO_2$[266] |
| DAS | 3.3 and 3.6 $\mu m$ | ··· | ··· | Hexane and ethanol vapor[147] |
| TDLAS | 2.566 $\mu m$ | 7 ppm | ··· | $C_2H_2$[141] |
| FTS | 3.7–4.05 $\mu m$ | ··· | 3 nm (resolution limit) | $N_2O$ spectrum reconstruction[160] |
| DCS | 2.6–4.1 $\mu m$ | ··· | 4.2 cm-1 (resolution) | Acetone (liquid phase)[267] |
| WMS | 3.291 $\mu m$ | 348 ppm | ··· | Methane[268] |

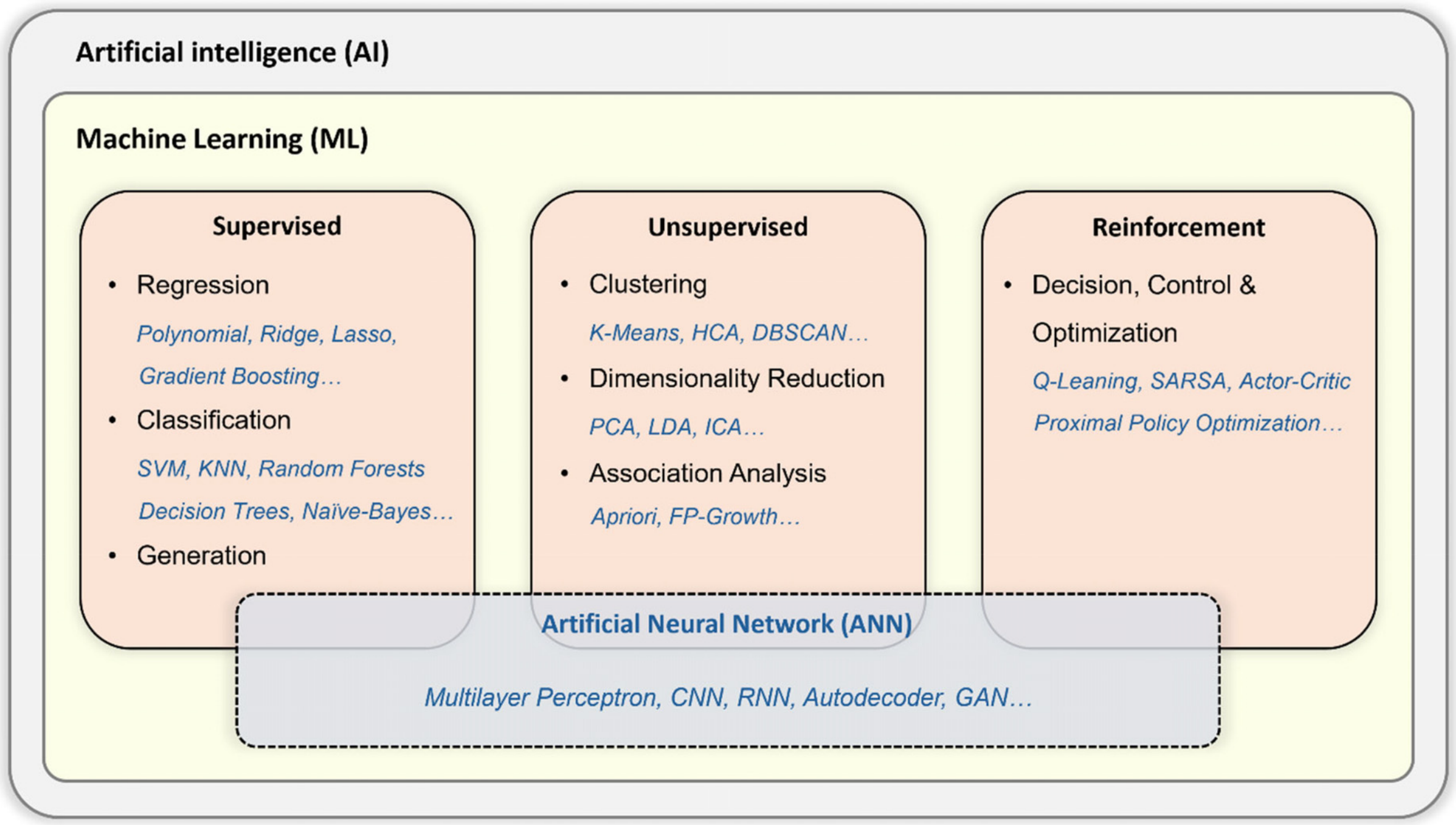

**FIG. 14.** Relationship between AI, ML, and NN, along with key algorithms and their primary applications. SVM, Support Vector Machine; KNN: K-Nearest Neighbors; HCA: Hierarchical Clustering Analysis; DBSCAN: Density-Based Spatial Clustering of Applications with Noise; PCA: Principal Component Analysis; LDA: Linear Discriminant Analysis; ICA: Independent Component Analysis; SARSA: State-Action-Reward-State-Action; CNN: Convolutional Neural Network; RNN: Recurrent Neural Network; GAN: Generative Adversarial Network.

Biosensors often generate time-dependent data, such as the continuous monitoring of physiological parameters, dynamic responses to stimuli, or sequential molecular interactions. The architecture of the chosen model is paramount, with considerations such as recurrent neural networks (RNNs) for sequential data. Zhang *et al.* used RNNs to quantify microRNAs with theory-guided supervision to improve classification metrics by 13.8%, achieving 98.5% accuracy in (let-7a) concentrations on cantilever biosensors and reducing false results.[282]

Support vector machines (SVM) are widely used in biosensor applications, particularly for classification tasks, because of their effectiveness in handling high-dimensional data and binary classification problems. Rong *et al.* used a multilayer perceptron (MLP) and an SVM to process signals from a COVID-19 optical-based detector on top of a nanoporous silicon material-based biosensor. This study qualitatively detected SARS-CoV-2 at concentrations as low as 1 TCID50/mL. Data distribution patterns were analyzed using t-distributed stochastic neighbor embedding (t-SNE).[283]

It is worth noting that most neural networks are electronic-based because traditional implementations leverage digital electronic components, providing fast and efficient computation. Electronic circuits are well-suited for representing and manipulating interconnected nodes and computations inherent in neural network architectures. Nevertheless, electronic neural networks often consume substantial power and exhibit restricted parallelism in information processing, resulting in potential bottlenecks.

In contrast to conventional electronic-based neural networks, optical neural networks (ONNs) represent a state-of-the-art advancement that exploits light for intricate computations,[285,286] inspired by intricate connections in the brain. ONNs aim to accelerate AI and machine-learning tasks by leveraging the inherent parallelism and swift nature of light. These networks ingeniously leverage the distinctive characteristics of light, including interference, diffraction, and polarization, for collective calculations, thereby markedly enhancing processing speed and capacity.[287] However, several unresolved issues must be addressed before fully harnessing the potential of optical neural networks (ONNs). For instance, although matrix multiplication is accomplished optically in ONNs, the subsequent application of activation functions still relies on electrical processes, thereby limiting the overall speed advantage of the ONNs. Another challenge is the absence of a fully optical feedback loop in ONNs, which hinders the operation of models such as recurrent neural networks, which require recurrent connections for sequential data processing. These challenges underscore the need for further advancements to exploit the capabilities of optical neural networks fully.

In the field of biosensing, ONNs emerged as a game-changing technology poised to reshape the narrative of biological molecule detection, analysis, and interpretation.[288–290] The fusion of ONNs with biosensing platforms offers several advantages. Primarily, ONNs exhibit heightened sensitivity and pinpoint accuracy, enabling the detection of laser modulation frequencies or analyte concentrations concealed within intricate biological samples. This precision is pivotal for early identification of ailments, personalized healthcare, and ecological monitoring. Moreover, ONNs' parallel processing of ONNs processes fast-tracks data scrutiny, leading to real-time monitoring and swift decision-making, particularly pivotal in scenarios such as point-of-care testing, where prompt outcomes are instrumental for timely intervention.

## VI. CONCLUSION

In the harmonious convergence of scientific ingenuity and technological prowess, the field of on-chip sensing has witnessed a profound transformation propelled by advancements in optical microring resonators, Mach–Zehnder interferometers, and photonic crystal waveguides. These miniature marvels exploit the intricate interplay between light and matter, heralding a new era of sensing capabilities across various domains. Through meticulous theoretical and experimental analysis, researchers have unlocked and are continuously exploring the latent potential of these platforms, unveiling their unprecedented sensitivity and selectivity. Future developments in on-chip optical sensing aim to enhance real-time responsiveness, reliability, and environmental robustness, making these sensors indispensable tools in dynamic and remote applications. Furthermore, breakthroughs in mid-infrared sensing will expand the scope of highly specific biochemical analyses considering their molecular fingerprinting capabilities. The evolution of on-chip spectroscopic sensing, encompassing absorption-based spectroscopy, Fourier Transform Spectrometers, and Wavelength Modulation Spectroscopy, has marked a pivotal advancement in the transformative potential across diverse applications. By leveraging photonic devices, these techniques offer compact and portable solutions for environmental monitoring, biomedical diagnostics, and other applications. While challenges persist in enhancing the interaction length and environmental robustness, recent developments have underscored the promise of on-chip spectroscopic sensing in revolutionizing scientific exploration and technological innovation. Anticipated breakthroughs in this domain will likely include the seamless integration of spectrometers with tunable light sources and detectors to achieve unparalleled precision and miniaturization, broadening accessibility to the next generation of spectroscopy systems. The integration of AI/ML algorithms with on-chip biosensors represents a significant leap forward, enabling enhanced sensitivity, specificity, and real-time monitoring capabilities. Through tailored AI/ML approaches, biosensors can efficiently analyze complex biological data, leading to rapid biomolecule detection and biomarker identification. The emergence of optical neural networks further accelerates biosensing tasks, promising heightened accuracy and real-time monitoring for personalized healthcare, disease detection, and environmental monitoring. The future fusion of AI/ML with advanced photonic platforms is expected to enable self-adaptive sensors, capable of optimizing performance autonomously and providing instantaneous insights for critical decision-making. Through continuous efforts and advanced innovations, on-chip optical sensing is steadily advancing, paving the way for a future where the possibilities of light illuminate our path of sustainable progress.

## ACKNOWLEDGMENTS

The authors acknowledge the support from NASA and DOE under contract numbers NASA-80NSSC22PB125, DE-SC0023917, and DE-SC0024015.

## AUTHOR DECLARATIONS

### Conflict of Interest

The authors have no conflicts to disclose.

### Author Contributions

**Sourabh Jain:** Conceptualization (equal); Formal analysis (equal); Funding acquisition (equal); Investigation (equal); Project administration (equal); Supervision (equal); Validation (equal); Visualization (equal); Writing – original draft (equal); Writing – review & editing (equal). **May Hlaing:** Conceptualization (equal); Investigation (equal); Writing – original draft (equal). **Kang-Chieh Fan:** Formal analysis (supporting); Writing – review & editing (equal). **Jason Midkiff:** Conceptualization (equal); Data curation (equal); Formal analysis (equal); Project administration (equal); Writing – review & editing (lead). **Shupeng Ning:** Formal analysis (supporting); Writing – review & editing (supporting). **Chenghao Feng:** Formal analysis (equal); Validation (equal); Writing – review & editing (supporting). **Po-yu Hsiao:** Writing – review & editing (equal). **Patrick Camp:** Validation (supporting); Visualization (supporting). **Ray T. Chen:** Conceptualization (equal); Formal analysis (equal); Funding acquisition (equal); Project administration (equal); Resources (equal); Supervision (equal); Validation (equal); Visualization (equal); Writing – review & editing (equal).

## DATA AVAILABILITY

The data that supports the findings of this study are available within the article.